\documentclass[12pt,english]{article}
\usepackage{lmodern}
\usepackage{times}
\usepackage[letterpaper]{geometry}
\usepackage{url}
\usepackage{float}
\usepackage{graphicx}
\usepackage{setspace}
\usepackage{amsmath}
\usepackage{multirow}
\usepackage{natbib}
\usepackage{algorithm}
\usepackage{color}
\usepackage{algorithmic}
\usepackage{subfig}
\newcommand{\beqn}{\vspace{-0.25cm}\begin{eqnarray*}}
\newcommand{\eeqn}{\end{eqnarray*}}
\newcommand{\bneqn}{\vspace{-0.25cm}\begin{eqnarray}}
\newcommand{\eneqn}{\vspace{-0.25cm}\end{eqnarray}}
\newtheorem{remark}{Remark}

\makeatletter

\providecommand{\tabularnewline}{\\}

\usepackage{pdflscape}

\makeatother
\author{Philip A. Ernst, James R. Thompson, and Yinsen Miao \\Department of Statistics, Rice University}
\title{Tukey's transformational ladder for portfolio management}
\begin{document}
\maketitle

\begin{abstract}
Over the past half-century, the empirical finance community has produced vast literature on the advantages of the equally weighted S\&P 500 portfolio as well as the often overlooked disadvantages of the market capitalization weighted Standard and Poor's (S\&P 500) portfolio (see \cite{Bloom}, \cite{Uppal}, \cite{Jacobs}, \cite{Treynor}). However, portfolio allocation based on Tukey's transformational ladde have, rather surprisingly, remained absent from the literature. In this work, we consider the S\&P 500 portfolio over the 1958-2015 time horizon weighted by Tukey's transformational ladder (\cite{Tukey2}):  $1/x^2,\,\, 1/x,\,\, 1/\sqrt{x},\,\, \text{log}(x),\,\, \sqrt{x},\,\, x,\,\, \text{and} \,\, x^2$, where $x$ is defined as the market capitalization weighted S\&P 500 portfolio. Accounting for dividends and transaction fees, we find that the 1/$x^2$ weighting strategy produces cumulative returns that significantly dominates all other portfolios, achieving a compound annual growth rate of 18\% over the 1958-2015 horizon. Our story is furthered by a startling phenomenon: both the cumulative and annual returns of the $1/x^2$ weighting strategy are superior to those of the $1/x$ weighting strategy, which are in turn superior to those of the 1/$\sqrt{x}$ weighted portfolio, and so forth, ending with the $x^2$ transformation, whose cumulative returns are the lowest of the seven transformations of Tukey's transformational ladder. The order of cumulative returns precisely follows that of Tukey's transformational ladder. To the best of our knowledge, we are the first to discover this phenomenon.
\end{abstract}

\newpage

\section{Introduction} \label{sec1}

For over half a century, the empirical finance community has extensively documented the advantages of the equally weighted Standard and Poor's (S\&P 500) portfolio as well as the often overlooked disadvantages of the market capitalization weighted S\&P 500 portfolio (see \cite{Bloom}, \cite{Uppal}, \cite{Jacobs}, \cite{Treynor}). In these works, novel statistical methodology has been created with the express purpose of analyzing alternative portfolio allocations. However, portfolio allocation based on the one of the most fundamental of statistical theories for data analysis, that of the seven transformations of John Tukey's transformational ladder (\cite{Tukey}, \cite{Tukey2}), has, to the best of our knowledge, been overlooked by this large and impressive literature. \\
\indent The motivation of the present paper is to infuse John Tukey's transformational ladder into the empirical finance literature. We consider the S\&P 500 portfolio from 1958-2015, and weight it by the entries of John Tukey's transformational ladder (\cite{Tukey2}):  $1/x^2,\,\, 1/x,\,\, 1/\sqrt{x},\,\, \text{log}(x),\,\, \sqrt{x},\,\, x,\,\, x^2$ (here, $x$ is the market capitalization weighted portfolio, which we henceforth abbreviate as ``MKC"). Consider a market capitalization portfolio named ``$x$'' with two equities, equity I and equity II. Let equity I account for 40\% of the total market capitalization of the portfolio and equity II account for 60\% of the total market capitalization of the portfolio. The first of John Tukey's transformations, the $1/x^2$ transformation, would assign weight

\beqn
\frac{1/.4^2}{1/.4^2+1/.6^2},
\eeqn
(approximately .69) to equity I and a weight of approximately .31 to equity II. This logic for re-weighting a portfolio with two stocks is then naturally extended to re-weighting a portfolio with 500 stocks. The data obtained from The Center for Research in Security Prices (CRSP) also includes data on dividends for each stock, which we include throughout our analysis. We rebalance the portfolio \textit{monthly} throughout the manuscript\footnote{For a justification of this rebalancing frequency, see Appendix B.}. We further assume transaction administrative fees of \$1 (in 2015 dollars) per trade and, additionally, a long-run average bid-ask spread of .1\% of the closing value of the stock. For example, if our portfolio buys (or sells) 50 shares of a given stock closing at \$100, transaction fees of \$1+ 50*(.1/2)= \$3.5 are incurred.\\
\indent This work presents two main findings. The first is that the 1/$x^2$ weighting strategy produces cumulative returns that significantly dominate all other portfolios, achieving a compound annual growth rate of 18\% from 1958-2015. The second is that the $1/x^2$ portfolio's compound annual growth rate is superior to the $1/x$ portfolio, which is in turn superior to 1/$\sqrt{x}$ portfolio, and so forth, ending with the $x^2$ transformation, whose cumulative returns are the lowest of the seven transformations of John Tukey's transformational ladder. That is, the order of cumulative returns \textit{precisely follows} that of Tukey's transformational ladder.\\
\indent Without further delay, we present our key findings. In Table \ref{mean2} we display the compound annual growth rate (in \%) of the equally weighted S\&P 500 portfolio (EQU) and the seven Tukey transformations of the S\&P 500, calculated from 1958-2015.
\begin{table}[H] 
\begin{onehalfspace}
\noindent \begin{centering}
\setlength{\tabcolsep}{12pt}
\begin{tabular}{c|c|c|c|c|c|c|c}
\hline
$1/x^2$ & $1/x$ & $1/\sqrt{x}$ & $\text{log}(x)$ & EQU & $\sqrt{x}$ & MKC & $x^2$ \tabularnewline 
\hline
18.00\% & 17.53\% & 15.23\% & 13.80\% & 13.32\% & 11.73\% & 10.43\% & 8.69\%\tabularnewline
\hline
\end{tabular}
\par\end{centering}
\vspace{.5 cm}
\caption{Compound annual growth rates (in \%) of the EQU and the seven Tukey transformational ladder portfolios, calculated from 1958-2015.}
\label{mean2}
\end{onehalfspace}
\end{table}

Cumulative returns alone are insufficient for analyzing investment performance. To this end, we present \textit{annual} returns for the eight portfolios under consideration. The mean annual returns, presented in Table \ref{mean1} below, are calculated by taking an arithmetic mean of the 58 annual returns (1958-2015) for each portfolio.
\begin{table}[H] 
\begin{onehalfspace}
\setlength{\tabcolsep}{12pt}
\noindent \begin{centering}
\begin{tabular}{c|c|c|c|c|c|c|c}
\hline
$1/x^2$ & $1/x$ & $1/\sqrt{x}$ & $\text{log}(x)$ & EQU & $\sqrt{x}$ & MKC & $x^2$ \tabularnewline
\hline
23.92\% & 20.35\% & 17.40\% & 15.62\% & 15.03\% & 13.18\% &  11.81\% & 10.25\%\tabularnewline
\hline
\end{tabular}
\par\end{centering}
\vspace{.5 cm}
\caption{The mean annual returns (in \%) of the EQU and the seven Tukey transformational ladder portfolios, calculated by taking an arithmetic mean of the 58 annual returns (1958-2015) for each portfolio.}
\label{mean1}
\end{onehalfspace}
\end{table}

\noindent The associated sample standard deviations are as follows
\begin{table}[H] 
\begin{onehalfspace}
\setlength{\tabcolsep}{12pt}
\noindent \begin{centering}
\begin{tabular}{c|c|c|c|c|c|c|c}
\hline
$1/x^2$ & $1/x$ & $1/\sqrt{x}$ & $\text{log}(x)$ & EQU & $\sqrt{x}$ & MKC & $x^2$ \tabularnewline
\hline
39.54\% & 26.44\% & 22.29\% & 20.01\% & 19.30\% & 17.52\% &  16.98\% & 18.05\%\tabularnewline
\hline
\end{tabular}
\par\end{centering}
\vspace{.5 cm}
\caption{Sample standard deviations of annual returns (in \%) of the EQU and the seven Tukey transformational ladder portfolios, calculated by taking an the sample standard deviation of the 58 annual returns (1958-2015) for each portfolio.}
\label{sd1}
\end{onehalfspace}
\end{table}
\noindent The Sharpe ratios, calculated using a risk free rate of 1.75\%, are

\begin{table}[H] 
\begin{onehalfspace}
\setlength{\tabcolsep}{12pt}
\noindent \begin{centering}

\begin{tabular}{c|c|c|c|c|c|c|c}
\hline
$1/x^2$ & $1/x$ & $1/\sqrt{x}$ & $\text{log}(x)$ & EQU & $\sqrt{x}$ & MKC & $x^2$ \tabularnewline
\hline
56.07\% & 70.35\% & 70.21\% & 69.31\%  & 68.81\% & 65.24\%  &  59.25\% & 47.09\% \tabularnewline
\hline
\end{tabular}
\par\end{centering}
\vspace{.5 cm}
\caption{Sharpe ratios of the eight portfolios under consideration.} \label{Sharpe1}
\end{onehalfspace}
\end{table}

Table \ref{mean2} shows that the compound annual growth rate of the $1/x^2$ portfolio, at 18.00\%, beats the market capitalization weighted portfolio's compound annual growth rate of 10.43\% by a factor of 1.73. Table \ref{mean1} displays that the arithmetic mean annual return of the $1/x^2$ weighted portfolio, at 23.92\%, beats the market capitalization weighted portfolio's arithmetic mean return of 11.81\% by a factor of 2.03. \\
\indent In his foreword to \cite{Bogle} (an articulate defense of the S\&P 500 Index Fund), Professor Paul Samuelson writes that ``Bogle's reasoned precepts can enable a few million of us savers to become in twenty years the envy of our suburban neighbors--while at the same time we have slept well in these eventful times.'' To use a strategy which is beaten by many others might not necessarily be a good thing. And yet, the S\&P 500 market cap weighted portfolio is probably used more than any other. Indeed, the tables above show there are superior alternatives, and the equally weighted portfolio is one of the many available.\\
\indent The remainder of this paper is organized as follows. Section \ref{litrev} surveys related literature, with a specific emphasis on on the ``small-firm effect'' phenomenon. Section \ref{sec2} provides an overview of the dataset and of the calculations employed. Section \ref{sec3} offers an analysis of cumulative returns of the portfolios. Section \ref{change} supplements Section \ref{sec3} and Section \ref{bootstrap} offers bootstrap simulations. Section \ref{sec4} offers an analysis of annual returns of the portfolios, Section \ref{sec8} calculates VaR and cVaR, and Section \ref{sec5} offers concluding remarks. Appendices A, B, C, and D supplement the main manuscript.

\section{Related Literature} \label{litrev}
The literature on the role of Tukey transforms in modern data analysis is vast (see \cite{Thomp1}, \cite{Thomp2}, \cite{Thomp3}, and \cite{Tukey} for a few of the many resources). Remarkably, and to the best of our knowledge, there has been no existing literature on the use of Tukey transforms \textit{for the purposes of portfolio management}. However, the ``small-firm effect,'' which refers to the superior performance of small-cap stocks relative to large-cap stocks, may easily be confused with the Tukey transformational ladder.\\
\indent We briefly review some of the seminal empirical findings in the small-firm effect literature. The small-firm effect was first introduced into the literature by \cite{Banz}, who empirically showed that during 1936-1975, the bottom quintile of stocks listed on the New York Stock Exchange (NYSE) achieved a .40\% excess risk-adjusted return over all remaining stocks. The field was greatly furthered by \cite{Fama}, who took a sample of stocks from NYSE, Amex, and Nasdaq over the period 1963-1990 and showed that the smallest 10\% of firms in their database outperformed the largest 10\% by .63\% per month. However, empirical studies conducted since \cite{Fama} have largely concluded that the size effect died out sometime in the early 1980s. The seminal work of \cite{Horo} shows no evidence of size effect over the 1979-1995 time horizon and the work of \cite{Hirsh} argues that the size effect disappeared in 1983. We refer the reader to \cite{van} for a more complete accounting of this literature. \\
\indent We wish to emphasize that our empirical results (presented in Section \ref{sec3} and thereafter) neither contradict nor support the small-firm effect hypothesis, and therefore that results concerning Tukey's transformational ladder for portfolio management must be viewed as their own distinct phenomena. At the present time, we do not have sufficient empirical evidence that the $1/x^2$ portfolio strategy does not ride on the size effect, and this matter will be investigated in future research.

\section{Data and Index Methodology} \label{sec2}
Our data is the S\&P 500 index from January 1958 to December 2015. The dataset was acquired from the Center for Research in Security Prices (CRSP)\footnote{\url{http://www.crsp.com}}. CRSP provides a reliable list of S\&P 500 index constituents, their respective daily stock prices, shares outstanding, dividends, and any ``key'' events, such as stock splits and acquisitions. The dataset is updated accordingly when a company joins or leaves the S\&P 500 constituent list. Further, the index returns are calculated for each portfolio according to the ``index return formula'' as documented by CRSP\footnote{\url{http://www.crsp.com/products/documentation/crsp-calculations}}. CRSP computes the return on an index ($R_{t}$) as the weighted
average of the returns for the individual securities in the index according to the following equation
\bneqn 
\label{CRSP}
R_{t}=\dfrac{\sum_{i}\omega_{i,t}\times r_{i,t}}{\sum_{i}\omega_{i,t}}\label{eq:1},
\eneqn
where $R_{t}$ is the index return, $\omega_{i,t}$ is the weight
of security $i$ at time $t$, and $r_{i,t}$ is the return of security
$i$ at time $t$.\\ 

\subsection{Calculations}
Throughout all calculations, we begin with an investment of \$100,000 in 1958 dollars. According to the Consumer Price Index (CPI) from Federal Reserve Bank of St. Louis\footnote{\url{https://fred.stlouisfed.org/series/CPIAUCNS}}, this is equivalent to approximately \$827,010.5 in 2015 dollars. Throughout all calculations, the transaction fees, which were documented in the second paragraph of Section \ref{sec1}, are discounted according to the CPI (for example, an administrative transaction fee of \$1 in 2015 is equivalent to 12.1 cents in 1958). All portfolios, with the exception of those in Appendix B, are rebalanced monthly and the transaction fees are subtracted from the portfolio total at market close on the first trading day of every month. Dividends are included in all calculations.

\subsection{Transaction fees}

Below we provide a table of transaction fees incurred by each of the eight portfolios under consideration over the 1958-2015 horizon. All numbers are discounted according to the CPI. The total transaction fees are much lower for the MKC and EQU portfolios than the $1/x^2$ and $1/x$ portfolios, as these require the most frequent rebalancing.

\begin{table}[H]
\footnotesize
\begin{onehalfspace}
\begin{centering}
\begin{tabular}{c|r@{\extracolsep{0pt}.}l|r@{\extracolsep{0pt}.}l|r@{\extracolsep{0pt}.}l|r@{\extracolsep{0pt}.}l|r@{\extracolsep{0pt}.}l|r@{\extracolsep{0pt}.}l|r@{\extracolsep{0pt}.}l|c}
\hline 
 & \multicolumn{2}{c|}{$1/x^{2}$} & \multicolumn{2}{c|}{$1/x$} & \multicolumn{2}{c|}{$1/\sqrt{x}$} & \multicolumn{2}{c|}{$\text{\ensuremath{\log\left(x\right)}}$} & \multicolumn{2}{c|}{EQU} & \multicolumn{2}{c|}{$\sqrt{x}$} & \multicolumn{2}{c|}{MKC} & $x^{2}$\tabularnewline
\hline 
Administration & 0&171 mil & 0&171 mil & 0&171 mil & 0&171 mil & 0&171 mil & 0&171 mil & 0&171 mil & 0.171 mil\tabularnewline
\hline 
Bid-ask Spread & 10&967 mil & 4&093 mil & 0&694 mil & 0&126 mil & 0&035 mil & 0&118 mil & 0&117 mil & 0.098 mil\tabularnewline
\hline 
Total & 11&138 mil & 4&264 mil & 0&865 mil & 0&297 mil & 0&206 mil & 0&289 mil & 0&288 mil & 0.269 mil\tabularnewline
\hline 
\end{tabular}
\par\end{centering}
\end{onehalfspace}
\vspace{.5 cm}
\caption{Administration fee (\$1 per trade) and bid-ask spread (0.1\% of the closing price per
stock) for each of the eight portfolios under consideration from 1958-2015. All portfolios are rebalanced monthly.}

\end{table}

\newpage
\section{Cumulative returns from 1958 to 2015} \label{sec3}
We present our first main finding below in Figure \ref{fig3}. Figure \ref{fig3} displays the cumulative returns calculated from 1958 to 2015 of the equally weighted S\&P 500 portfolio (EQU) and the seven portfolios given by the Tukey transformations ($1/x^2,\,\, 1/x,\,\, 1/\sqrt{x},\,\, \text{log}(x),\,\, \sqrt{x},\,\,x,\,\,x^2$), where $x$ is the market capitalization weighted portfolio. The calculation assumes that \$100,000 (in 1958 dollars) is invested in each of portfolio on 1/2/58 and is left to grow until 12/31/15. \textit{All dividends and transaction fees are taken into account, here and in every figure and table produced in this work}.

\begin{figure}[H]
\noindent \centering{}\includegraphics[width=16.8cm]{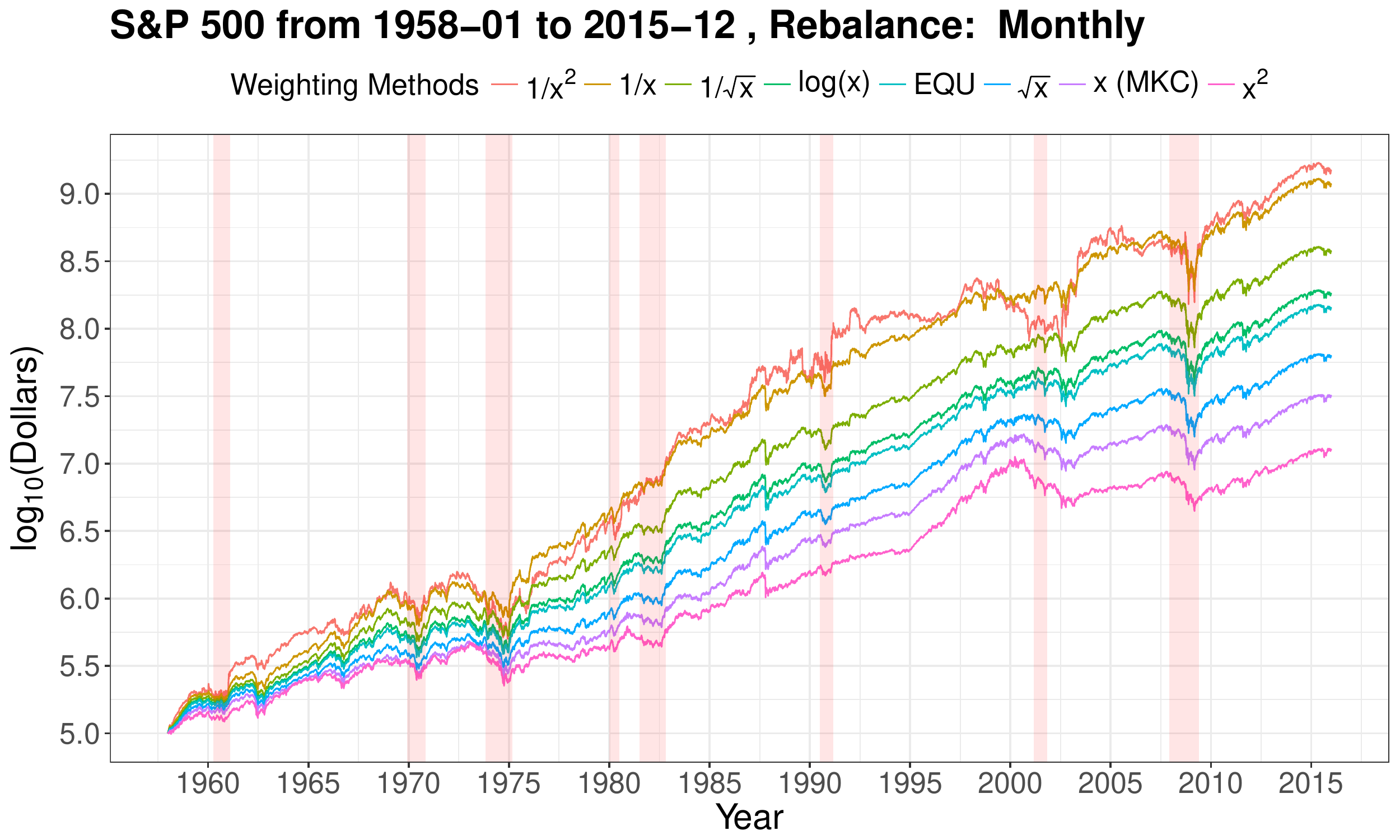}\caption{Cumulative $\text{log}_{10}$ returns (from 1958-2015) for the EQU portfolio and the seven Tukey transformational ladder portfolios. The calculation assumes that \$100,000 is invested on 1/2/58 and is left to grow until 12/31/15.} 
\label{fig3}
\end{figure}
Figure \ref{fig3} shows significant changes in portfolio returns over the 1958-2015 time horizon. Across all eight portfolios, the following macroeconomic events are well pronounced: the bear market from November 1968 to May 1970 (high inflation and the Vietnam war), the January 1973-October 1974 bear market (OPEC embargo), the 1982-1987 bull market, the ``Black Monday'' crash of October 1987, the 1988-2000 bull market, the dot-com bubble burst from March 2000 to October 2002, the most recent financial crisis of October 2007 to March 2009, and the most recent bull market which began in March 2009. \\
\indent The cumulative returns on 12/31/15 displayed in Figure \ref{fig3} are reproduced in Table \ref{table2} below.\\

\begin{table}[H]
\begin{onehalfspace}
\footnotesize
\centering{}%
\begin{tabular}{rrrrrrrr}
\hline 
$1/x^{2}$ & $1/x$  & $1/\sqrt{x}$  & $\log\left(x\right)$  & EQU  & $\sqrt{x}$  & $x$ & $x^{2}$\tabularnewline
\hline 
\$ 1.477 bil &\$ 1.169 bil &\$ 372.539 mil &\$ 180.263 mil&\$ 141.373 mil& \$ 62.217 mil&\$ 31.516 mil &\$ 12.544 mil \tabularnewline
\hline 
\end{tabular}\caption{Cumulative values on 12/31/15 for EQU and all seven Tukey transformational ladder portfolios.} \label{table2}
\end{onehalfspace}
\end{table}
\normalsize

\subsection{Discussion of Table \ref{table2}} \label{sec41}
Our first main finding is exhibited by Table \ref{table2}, which shows that investing \$100,000 in January 1958 in the  $1/x^2$ portfolio yields a cumulative value of \$1.477 billion in December 2015. As such, the $1/x^2$ portfolio's value on 12/31/15 remarkably dominates all other seven portfolios by a \textit{substantial} margin; in particular, it exceeds the market capitalization's cumulative value of \$31.516 million by a factor of 46.865! The dominance of the $1/x^2$ weighted portfolio cumulative return will be explored on theoretical grounds in a future paper and as such is beyond the scope of the present work. For the purposes of this paper, we favor an intuitive explanation. The $1/x^2$ portfolio assigns the large majority of its weight to the lowest cap stocks of the S\&P 500, very little weight to the larger cap stocks of the S\&P 500, and negligible weight to the largest cap stocks of the S\&P 500. Consequently, the portfolio reaps the benefits from the ``smaller cap stocks'' of the S\&P 500, the latter of which are more volatile and may present more opportunity for growth.\\
\indent Our second main finding from Table \ref{table2} is that the cumulative values of the portfolios follow the precise order of Tukey's transformational ladder. Namely, the cumulative value of the $1/x^2$ portfolio is largest, followed by the $1/x$, $1/\sqrt{x}$, $\text{log}(x)$, $\sqrt{x}$, and $x$ (MKC) portfolios, and ending with the $x^2$ portfolio. To the best of our knowledge, we are the first to discover this phenomenon.

\begin{remark}
The rabbis of the Babylonian Talmud are often credited to be the first to give explicit advice on wealth allocation. In the fourth century, Rabbi Isaac Bar Aha, wrote that ``one should always divide his wealth into three parts: a third in land, a third in merchandise, and a third ready to hand.''\footnote{The Babylonian Talmud, tractate Baba Mezi'a, volume 42a.} Unlike Rabbi Isaac Bar Aha, perhaps the late John Tukey may have unknowingly offered suggestions for wealth management.
\end{remark}

\section{The Tukey transformational ladder for alternate time horizons} \label{change}
Figure \ref{fig3} in Section \ref{sec3} shows that the portfolio returns precisely follow the Tukey transformational ladder over the 1958-2015 time horizon.  A natural line of inquiry is to determine whether the portfolio returns precisely follow the Tukey transformational ladder for other time horizons. We thus proceed to calculate the cumulative returns of the for the EQU and the seven Tukey transformations for four additional time periods: 1970-2015, 1980-2015, 1990-2015, and 2000-2015. The portfolio returns precisely follow the order of the Tukey transformational ladder over these additional four time periods.\\
\indent We first consider the time horizon 1970-2015. We invest \$132,168  on 1/2/70 (the equivalent of \$100,000 in 1958 dollars) and let the portfolios grow until 12/31/15. The following cumulative portfolio values are presented in Figure \ref{fig9} below.

\begin{figure}[H]
\noindent \begin{centering}
\includegraphics[width=16.6cm]{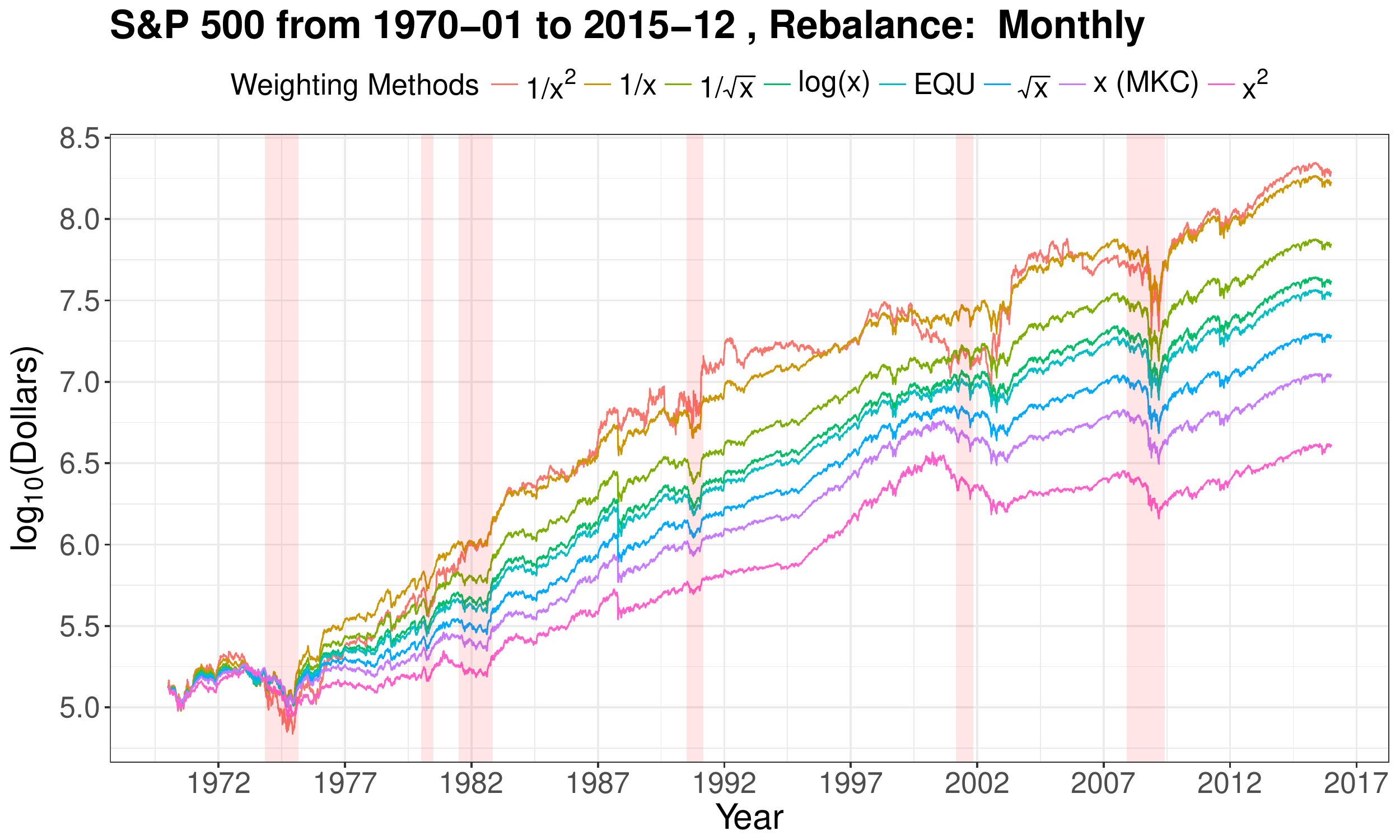}
\par\end{centering}

\caption{Cumulative $\text{log}_{10}$ returns (from 1970-2015) for the EQU portfolio and the seven Tukey transformational ladder portfolios. The calculation assumes that \$132,168  is invested on 1/2/70 and is left to grow until 12/31/15.} \label{fig9}
\end{figure}
\noindent The cumulative returns displayed in Figure \ref{fig9} are reproduced in Table \ref{table9} below.

\begin{table}[H]
\begin{onehalfspace}
\footnotesize
\noindent \centering{}%
\begin{tabular}{rrrrrrrr}
\hline 
$1/x^{2}$ & $1/x$ & $\sqrt{x}$ & $\log\left(x\right)$  & EQU & $\sqrt{x}$ & $x$ & $x^{2}$\tabularnewline
\hline 
\$192.505 mil  & \$166.178 mil & \$69.009 mil  & \$40.900 mil & \$34.410 mil  & \$18.964  mil  & \$10.904 mil & \$4.028 mil \tabularnewline
\hline 
\end{tabular}\caption{The cumulative returns for the EQU portfolio and the Tukey transformational ladder portfolios. The caclulation assumes that \$132,168 is invested on 1/2/70 and is left to grow until 12/31/15.} \label{table9}
\end{onehalfspace}
\end{table}
The graphs and tables for the 1980-2015, 1990-2015, and 2000-2015 time horizons appear in Appendix D.
\section{Bootstrap} \label{bootstrap}
We bootstrap each portfolio to obtain confidence intervals for each portfolio's cumulative returns. Section \ref{sub61} shows that the mean returns of the bootstrap distributions for random $N$ precisely follow the ``modified'' Tukey transformational ladder in the following sense: if one omits the $1/x^2$ portfolio from consideration, the bootstrapped means precisely follow the remainder of the Tukey transformational ladder (the $1/x$ transformation has the highest bootstrapped sample mean, followed by $1/\sqrt{x}$, $\text{log}(x)$, $\sqrt{x}, x$, and culminating with $x^2$). 

\subsection{Bootstrap for Random $N$} \label{sub61}
\begin{figure}[H]
\noindent \centering{}\includegraphics[width=18cm]{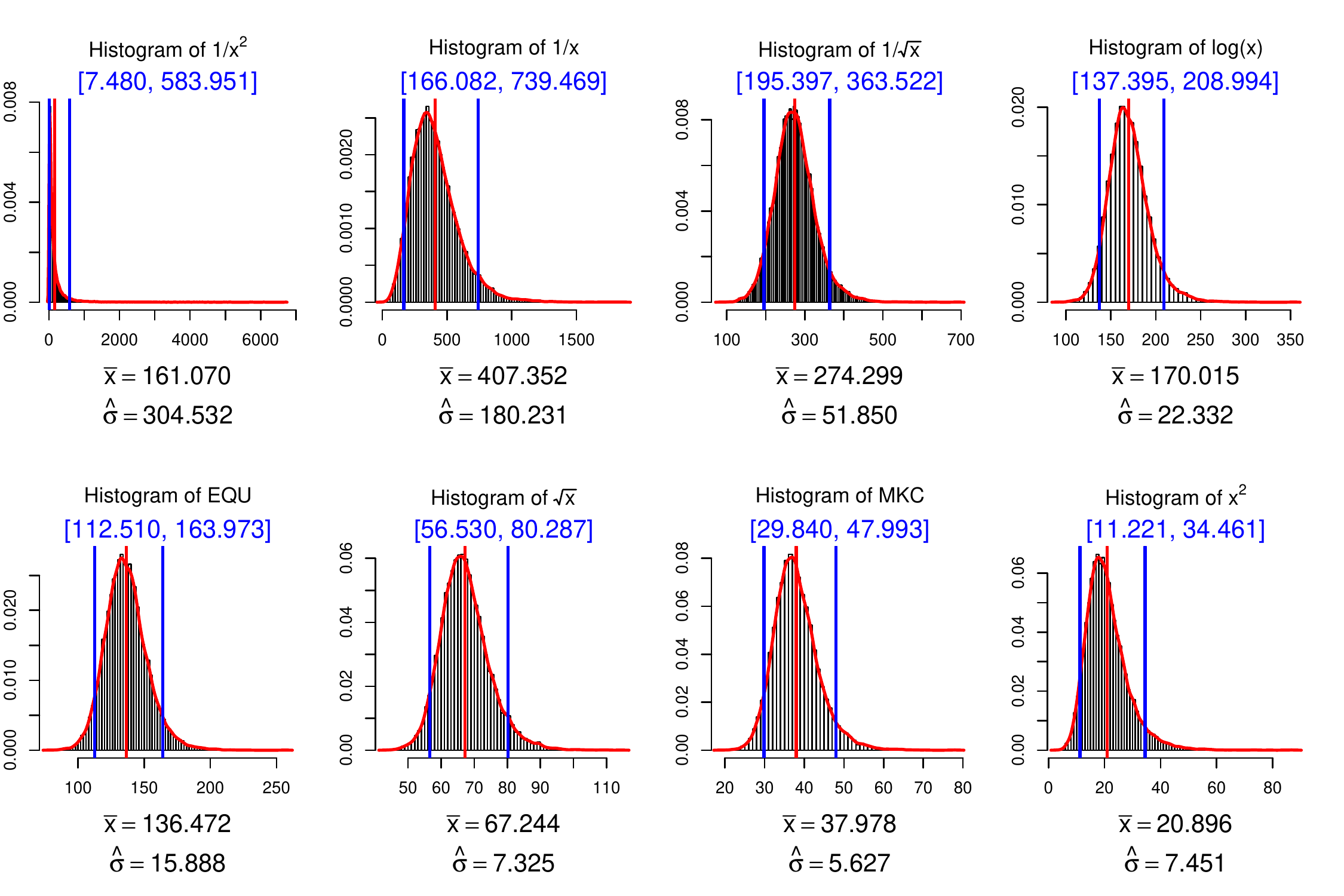}\caption{The bootstrap distribution produced using Algorithm~\ref{algo:1}. The estimated mean $\bar{x}$, standard deviation
$\hat{\sigma}$, and 95\% confidence intervals are displayed. All numbers are in million USD. \label{fig:1}}
\end{figure}

\indent We conduct a simple bootstrap as illustrated below in Algorithm \ref{algo:1}. First, we uniformly choose the number of stocks $N$ from the
sample space $\Omega=\left\{ 100,101,\ldots,500\right\} $. Second, we sample $N$ stocks with replacement from all listed
stocks in S\&P 500 from 1/2/58 to 12/31/15. We proceed to calculate the subsequent
daily return using CRSP's return on index formula\\
\[
R_{t}=\dfrac{\sum_{i}\omega_{i,t}\times r_{i,t}}{\sum_{i}\omega_{i,t}},
\]
where $R_{t}$ is the portfolio return on day $t$, $\omega_{i,t}$
is the weight of security $i$ on day $t$, and $r_{i,t}$ is the
return of the security $i$ on day $t$. The variable $\omega_{i,t}$ is computed
using a Tukey transformation of the market capitalization rate on day $t-1$.
We then compute the cumulative returns using these daily returns and repeat the above process 20,000 times. The resulting bootstrap plots are presented in Figure \ref{fig:1}, and the units are in million USD.  The key idea is that the mean returns of the bootstrap distributions precisely follow the ``modified'' Tukey transformational ladder in the following sense: if one omits the $1/x^2$ portfolio from consideration, the bootstrapped means precisely follow the remainder of the Tukey transformational ladder (the $1/x$ transformation has the highest bootstrapped sample mean, followed by $1/\sqrt{x}$, $\text{log}(x)$, $\sqrt{x}, x$, and finally culminating with $x^2$).

\newpage
\noindent Higher resolution plots of the $1/x^2$ and $1/x$ bootstrap distributions are found below.
\begin{figure}[H]

\centering
    \subfloat[$1/x^2$]{{\includegraphics[width=8cm]{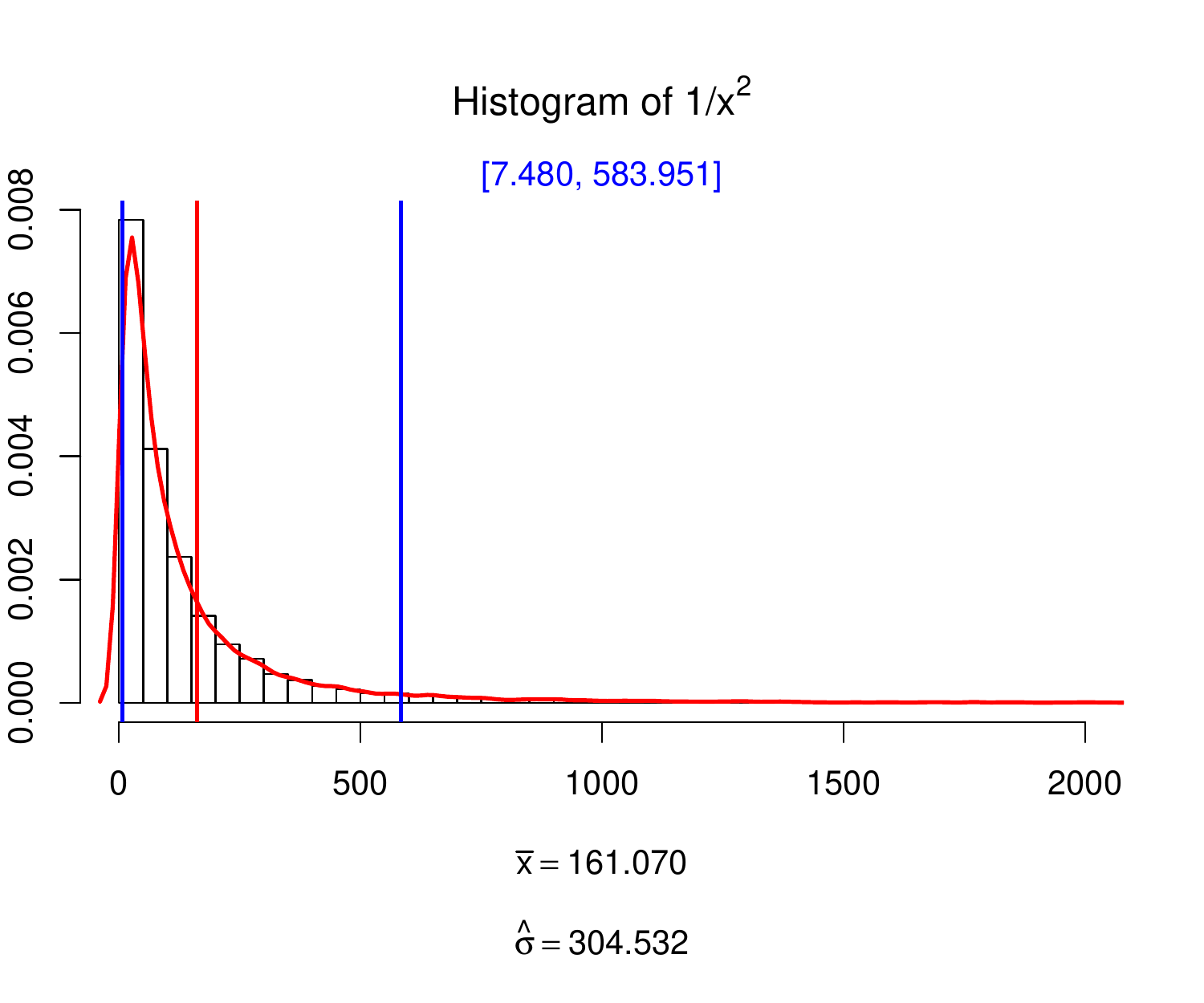} }}%
    \qquad
    \subfloat[$1/x$]{{\includegraphics[width=8cm]{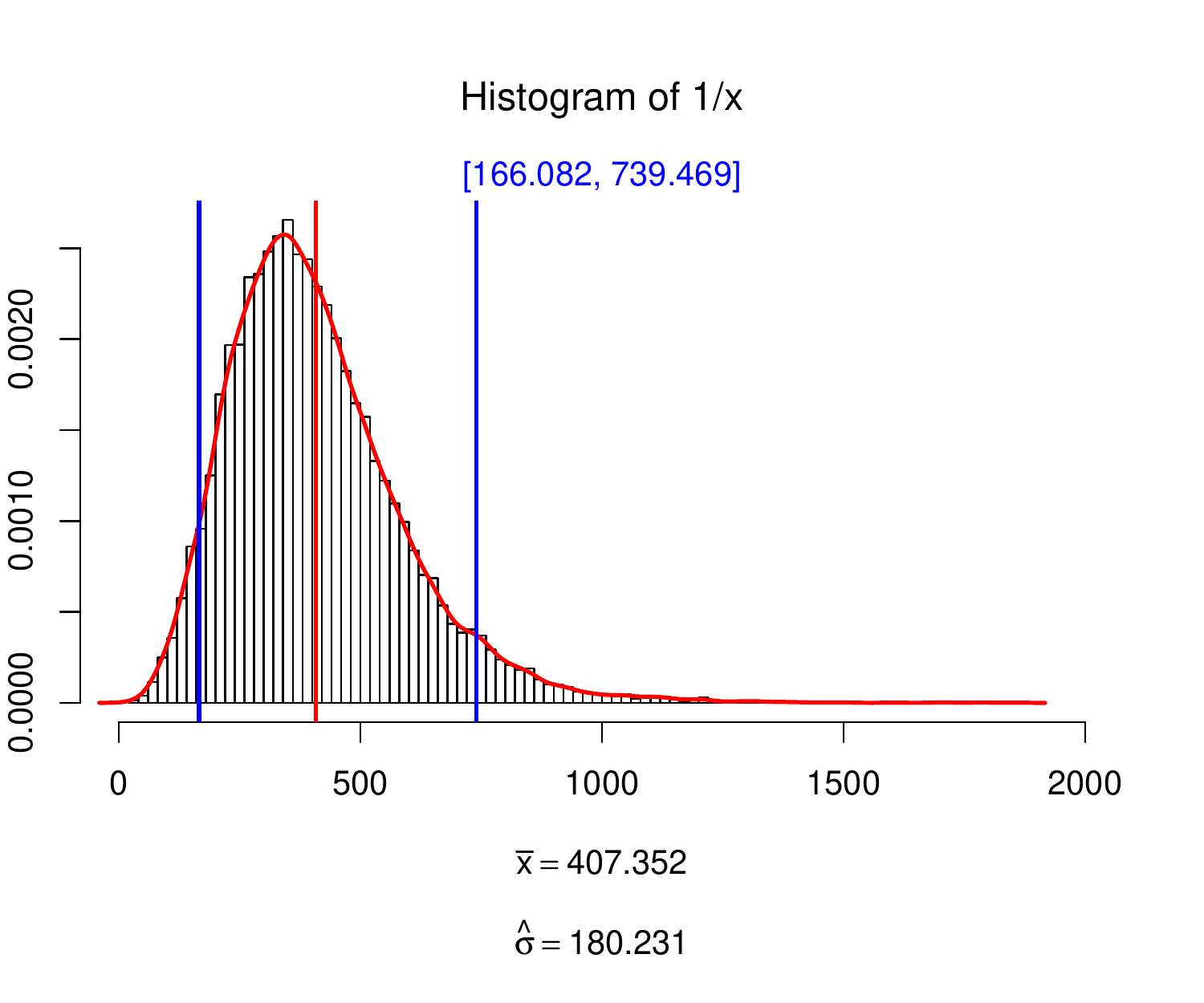} }}%
\caption{$1/x^2$ and $1/x$  bootstrap distributions.\label{fig:2}}

\end{figure}

\begin{algorithm}[H]
\begin{algorithmic}

\FOR{$itr$ in $1,\cdots, 10000$}
	\STATE Sample $N$ from $\Omega=\left\{ 100,101,\cdots,500\right\} $.
	\STATE Sample $N$ stocks from the S\&P 500 list randomly with replacement.
		\FOR{$t$ from 1958/01/03 to 2015/12/31}
			\IF{$k$ stocks are deleted from our selected portfolio on day $t$.}
				\STATE Randomly select the other $k$ remaining stocks in S\&P 500 on day $t$ with replacement.
			\ENDIF
			\STATE Compute daily return $R_{t}$ for day $t$
				\[ R_{t}=\dfrac{\sum_{i}\omega_{i,t}\times 				r_{i,t}}{\sum_{i}\omega_{i,t}} \]
		\ENDFOR
	\STATE Compute the cumulative return for iteration $itr$.
		\[ CR_{itr}=10^5\times\prod_{t=1}\left(1+R_{t}\right) \] 
\ENDFOR
\end{algorithmic}
\caption{Bootstrap Sampling}
\label{algo:1}
\end{algorithm}

\subsection{Bootstrap for Fixed $N$} \label{sec62}
We now modify Algorithm 1 to conduct a bootstrap for fixed $N$. Below we produce the bootstrapped simulations for $N=10$ for the EQU portfolio and the seven portfolios of the Tukey transformational ladder for the 1/2/58-12/31/15 horizon. The following sample statistics are reported: the 1st percentile, the 5th percentile, the median, the mean, the 95th percentile, and the 99th percentile. The blue lines in each plot denote the 5th and 95th percentiles of the bootstrapped distribution.

\begin{figure}[H]
\noindent \begin{centering}
\includegraphics[width=16cm]{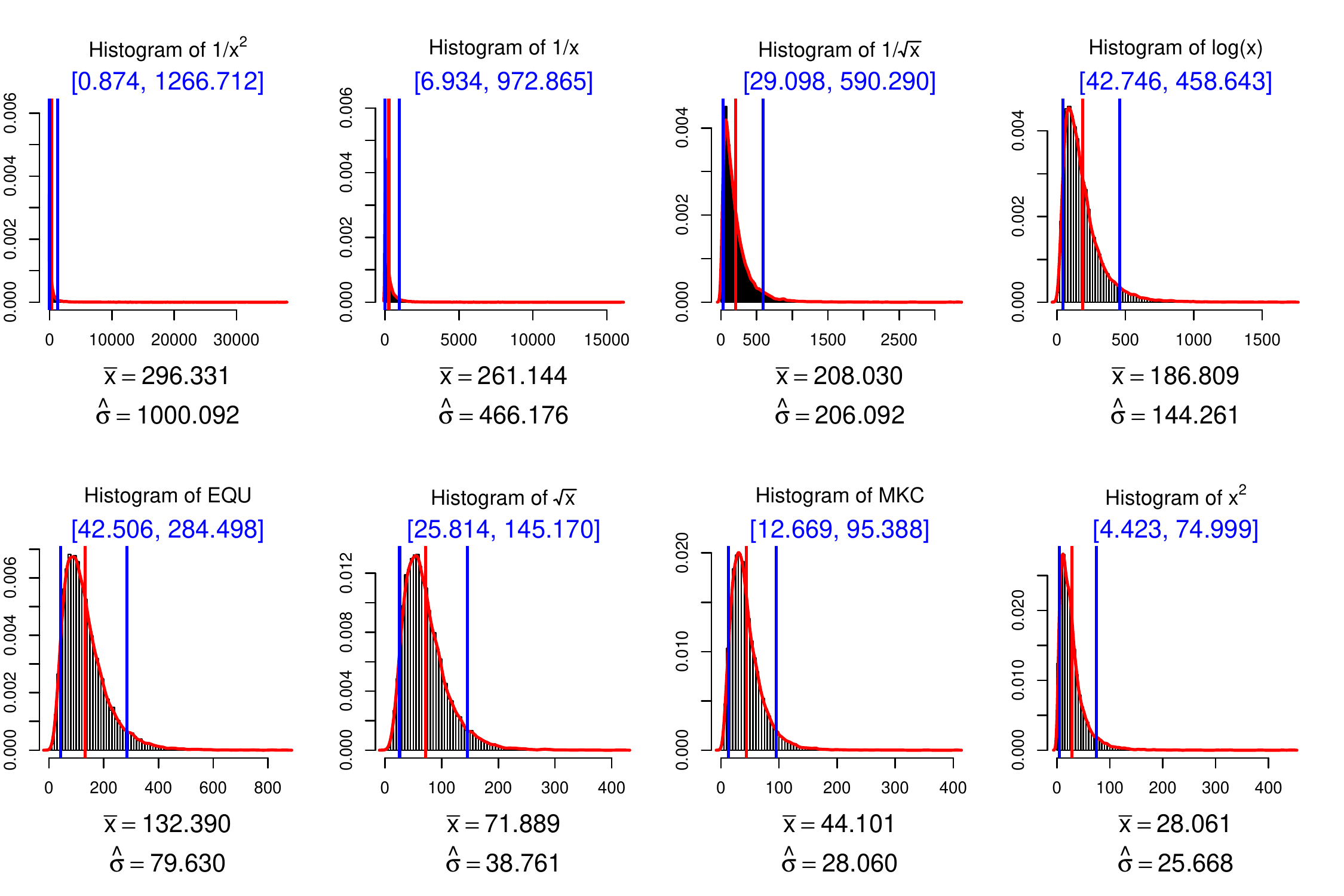}\caption{The bootstrap distributions for $N=10$.}
\par\end{centering}
\end{figure}

\begin{table}[H]
\begin{onehalfspace}
\noindent \begin{centering}
\begin{tabular}{rrrrrrrr}
\hline 
 & mean  & median  & sd  & q5th  & q95th  & q1th & q99th\tabularnewline
\hline 
$1/x^{2}$ & 296.331 & 50.890 & 1000.092 & 0.874 & 1266.712 & 0.135 & 3970.265 \tabularnewline 
$1/x$ & 261.144 & 113.067 & 466.176 & 6.934 & 972.865 & 1.905 & 2158.401 \tabularnewline 
$1/\sqrt{x}$ & 208.030 & 147.852 & 206.092 & 29.098 & 590.290 & 13.791 & 993.413 \tabularnewline 
 $\log\left(x\right)$ & 186.809 & 149.495 & 144.261 & 42.746 & 458.643 & 24.520 & 716.204 \tabularnewline 
EQU & 132.390 & 114.843 & 79.630 & 42.506 & 284.498 & 27.477 & 403.269 \tabularnewline 
$\sqrt{x}$ & 71.889 & 63.964 & 38.761 & 25.814 & 145.170 & 16.674 & 200.863 \tabularnewline 
$x$ & 44.101 & 37.897 & 28.060 & 12.669 & 95.388 & 7.260 & 141.100 \tabularnewline 
$x^{2}$ & 28.061 & 21.734 & 25.668 & 4.423 & 74.999 & 1.997 & 122.697 \tabularnewline 
\hline 
\end{tabular}
\par\end{centering}
\end{onehalfspace}
\caption{Sample statistics for the cumulative return on 12/31/15 for $N=10$, calculated from 20,000 simulations. All numbers are in million USD.}
\end{table}

In the above results for $N=10$, the mean of the $1/x^2$ portfolio is highest, and the bootstrapped sample means precisely follow the order of the Tukey transformational ladder. The same holds true for the bootstrapped sample means for $N=20$ (see Appendix C). However, for $N=50$ and higher (see Appendix C), the $1/x$ bootstraped portfolio posts the highest sample mean. For these higher values of $N$, the bootstrapped means follow the ``modified'' Tukey transformational ladder. For $N=200,300,400,500$, the sample mean for the $1/x^2$ bootstrapped distribution falls above EQU, but below that of $\text{log}(x)$. We conclude that the $1/x^2$ transformation is not robust for fixed $N$ for large values of $N$.

\section{Annual rates of return} \label{sec4}
\indent Below we display the key findings from Tables \ref{table4} and \ref{table5} in Appendix A. Table \ref{mean12} gives the mean annual returns (in \%) of the EQU and the seven Tukey transformational ladder portfolios, calculated by taking an arithmetic mean of the 58 annual returns (1958-2015) for each portfolio.

\begin{table}[H] 
\begin{onehalfspace}
\setlength{\tabcolsep}{12pt}
\noindent \begin{centering}
\begin{tabular}{c|c|c|c|c|c|c|c}
\hline
$1/x^2$ & $1/x$ & $1/\sqrt{x}$ & $\text{log}(x)$ & EQU & $\sqrt{x}$ & MKC & $x^2$ \tabularnewline
\hline
23.92\% & 20.35\% & 17.40\% & 15.62\% & 15.03\% & 13.18\% &  11.81\% & 10.25\%\tabularnewline
\hline
\end{tabular}
\par\end{centering}
\vspace{.5 cm}
\caption{The mean annual returns (in \%) of the EQU and the seven Tukey transformational ladder portfolios, calculated by taking an arithmetic mean of the 58 annual returns (1958-2015) for each portfolio.}
\label{mean12}
\end{onehalfspace}
\end{table}
\noindent The associated sample standard deviations are as follows:

\begin{table}[H] 
\begin{onehalfspace}
\setlength{\tabcolsep}{12pt}
\noindent \begin{centering}
\begin{tabular}{c|c|c|c|c|c|c|c}
\hline
$1/x^2$ & $1/x$ & $1/\sqrt{x}$ & $\text{log}(x)$ & EQU & $\sqrt{x}$ & MKC & $x^2$ \tabularnewline
\hline
39.54\% & 26.44\% & 22.29\% & 20.01\% & 19.30\% & 17.52\% &  16.98\% & 18.05\%\tabularnewline
\hline
\end{tabular}
\par\end{centering}
\vspace{.5 cm}
\caption{Sample standard deviations of annual returns (in \%) of the EQU and the seven Tukey transformational ladder portfolios, calculated by taking an the sample standard deviation of the 58 annual returns (1958-2015) for each portfolio.}
\label{sd12}
\end{onehalfspace}
\end{table}
\noindent The associated Sharpe ratios, used a risk free rate of 1.75\%, are:

\begin{table}[H] 
\begin{onehalfspace}
\setlength{\tabcolsep}{12pt}
\noindent \begin{centering}
\begin{tabular}{c|c|c|c|c|c|c|c}
\hline
$1/x^2$ & $1/x$ & $1/\sqrt{x}$ & $\text{log}(x)$ & EQU & $\sqrt{x}$ & MKC & $x^2$ \tabularnewline
\hline
56.07\% & 70.35\% & 70.21\% & 69.31\%  & 68.81\% & 65.24\%  &  59.25\% & 47.09\% \tabularnewline
\hline
\end{tabular}
\par\end{centering}
\vspace{.5 cm}
\caption{Sharpe ratios of the eight portfolios under consideration.} \label{sharpe1}
\label{Sharpe12}
\end{onehalfspace}
\end{table}

The cumulative returns of the $1/x^2$ portfolio, as presented in Table \ref{table2}, in addition to the average annual returns presented above in Table \ref{Sharpe12}, indeed make it a tempting strategy for investment professionals and hedge funds. However, due to both its large standard deviation (see Table \ref{sd12}) and extremely high values of VaR and cVaR (see Section \ref{sec8}), the $1/x^2$ portfolio presents enormous risk, even for investors with long-term horizons. Investors should instead consider the $1/x$ weighted portfolio, which posts the highest Sharpe ratio of the eight portfolios under consideration and enjoys more moderate values of VaR and cVaR than those of $1/x^2$ (see Section \ref{sec8}).\\
\indent  Finally, it should be noted that the $1/x^2$ and the $1/x$ strategies are contrarian strategies, as they buy declining equities, whereas the $x^2$ strategy, being that it buys rising equities, represents a momentum strategy. For further discussion on the merits of both momentum and contrarian strategies, we refer the reader to \cite{Chan}, \cite{Goetz}, \cite{Yao}, and \cite{Franck}. Finally, we wish to emphasize that the $1/x^2$ and $1/x$ strategies are both strategies that could only be executed by niche players. It should be noted that if sufficiently many large players sought to implement the $1/x^2$ and $1/x$ portfolios, the financial markets would likely no longer reward the niche players utilizing the $1/x^2$ and $1/x$ strategies.

\subsection{Large annual returns}
\indent Tables \ref{table4} and \ref{table5} (see Appendix A) post considerably large annual returns (in absolute value). This is not only so for the $1/x^2$ transformation, but for the other Tukey transformations as well. To justify these large returns, we produce Table \ref{fig4}, which reports the number of S\&P 500 constituents in each calendar year which grew by 50\% or more. For example, in 1976, 88 companies grew by at least 50\% and 14 companies grew by at least 100 \%. Therefore, it is within reason to calculate a 1976 annual return for the $1/x^2$ portfolio of 98.00\%. Note that in 2009, 143 companies grew by at least 50\%, 45 companies grew by at least 100\%, and 14 companies grew by at least 200\%.

\begin{table}[h!]
\noindent \begin{centering}
\begin{tabular}{rrrr|rrrr|rrrr}
\hline 
Year & 50\%  & 100\%  & 200\%  & Year & 50\%  & 100\%  & 200\%  & Year & 50\%  & 100\%  & 200\% \tabularnewline
\hline 
1958  & 160  & 26  & 6  & 1977  & 8  & 4  & 0  & 1996  & 40  & 6  & 0 \tabularnewline
1959  & 35  & 8  & 0  & 1978  & 27  & 6  & 3  & 1997  & 104  & 13  & 3 \tabularnewline
1960  & 10  & 0  & 0  & 1979  & 86  & 19  & 4  & 1998  & 74  & 19  & 5 \tabularnewline
1961  & 73  & 10  & 0  & 1980  & 95  & 26  & 3  & 1999  & 85  & 34  & 10 \tabularnewline
1962  & 1  & 0  & 0  & 1981  & 25  & 3  & 1  & 2000  & 79  & 15  & 3 \tabularnewline
1963  & 49  & 7  & 1  & 1982  & 111  & 20  & 5  & 2001  & 32  & 11  & 0 \tabularnewline
1964  & 34  & 6  & 0  & 1983  & 78  & 14  & 1  & 2002  & 3  & 0  & 0 \tabularnewline
1965  & 78  & 15  & 2  & 1984  & 19  & 4  & 2  & 2003  & 131  & 34  & 7 \tabularnewline
1966  & 4  & 1  & 0  & 1985  & 91  & 8  & 1  & 2004  & 40  & 3  & 2 \tabularnewline
1967  & 115  & 30  & 4  & 1986  & 49  & 7  & 0  & 2005  & 27  & 4  & 0 \tabularnewline
1968  & 66  & 10  & 1  & 1987  & 38  & 9  & 2  & 2006  & 23  & 2  & 0 \tabularnewline
1969  & 3  & 2  & 0  & 1988  & 49  & 14  & 1  & 2007  & 35  & 11  & 3 \tabularnewline
1970  & 19  & 0  & 0  & 1989  & 88  & 7  & 1  & 2008  & 3  & 3  & 3 \tabularnewline
1971  & 53  & 6  & 0  & 1990  & 4  & 0  & 0  & 2009  & 143  & 45  & 14 \tabularnewline
1972  & 31  & 3  & 0  & 1991  & 131  & 29  & 4  & 2010  & 56  & 5  & 2 \tabularnewline
1973  & 19  & 7  & 0  & 1992  & 42  & 10  & 1  & 2011  & 11  & 1  & 0 \tabularnewline
1974  & 10  & 5  & 1  & 1993  & 44  & 10  & 0  & 2012  & 33  & 5  & 0 \tabularnewline
1975  & 187  & 50  & 3  & 1994  & 17  & 2  & 0  & 2013  & 121  & 12  & 5 \tabularnewline
1976  & 88  & 14  & 0  & 1995  & 96  & 10  & 3  & 2014  & 13  & 2  & 0 \tabularnewline
 &  &  &  &  &  &  &  & 2015 & 7  & 3  & 1 \tabularnewline
\hline 
\end{tabular}
\par\end{centering}
\caption{Number of S\&P 500 constituents whose stock price increased by at least 50\%, 100\% or 200\%.}
\label{fig4}
\end{table}
A striking feature of the $1/x^2$ portfolio is that, despite its larger volatility, it fared quite well (particularly in comparison with the market capitalization weighted portfolio) during the market crashes of 2001 and 2008. Table \ref{table5} (see Appendix) shows that in 2001, $1/x^2$ gained 6.30\%, whereas MKC lost approximately 11.82\%. In 2008, $1/x^2$ lost approximately 33.54\%, whereas MKC lost approximately 35.25 \%. It is also worthy to note that of the 58 years from 1958-2015, the $1/x^2$ portfolio posts 18 years with negative returns and the $1/x$ portfolio posts 12 years with negative returns. The latter figure is only slightly more than EQU, which posts 11 years with negative returns, and slightly less than MKC, which posts 13 years of negative returns.

\section{Calculation of VaR and cVaR} \label{sec8}

An analysis of investment performance based on the first and second moments alone is insufficient. In this vein, we calculate the VaR (value at risk) at 5\% for each of the portfolios in Table \ref{table40} below and the expected shortfall (cVaR) at 5\%  for each of the portfolios in Table \ref{table41} below. For additional measures of potential shortfall, we refer the reader to \cite{Kadan}.

\begin{table}[H]
\noindent \begin{centering}
\begin{tabular}{rccc}
\hline 
&Annual VaR & Monthly VaR & Daily VaR \tabularnewline
\hline 
$1/x^{2}$ & -33.96  & -10.11 & -2.62 \tabularnewline
$1/x$ & -16.60  & -7.38 & -1.61  \tabularnewline
$1/\sqrt{x}$ & -18.65  & -7.13 &-1.49\tabularnewline
$\log\left(x\right)$ & -18.91  & -6.78 &-1.45  \tabularnewline
EQU  & -17.98  & -6.63 & -1.46  \tabularnewline
$\sqrt{x}$ & -17.43  & -6.43 & -1.44   \tabularnewline
MKC  & -15.98  & -6.40 & -1.48  \tabularnewline
$x^{2}$ &  -24.23  & -6.72 & -1.62  \tabularnewline
\hline 
\end{tabular}
\par\end{centering}
\caption{Annual VaR, Monthly VaR, and Daily VaR for the EQU portfolio and the seven Tukey transformational ladder portfolios. All numbers are expressed in \%.} \label{table40}
\end{table}
Given the highly skewed (to the right) distributions of $1/x^2$ in Section \ref{bootstrap} as well as in Appendix C, it is not surprising to see large (negative) values for VaR for the $1/x^2$ strategy at 5\%. The values of VaR for $1/x$ (which posts the highest Sharpe ratio of the eight portfolios under consideration) are much closer to the values of VaR for the EQU and MKC portfolios than the $1/x^2$ portfolio. This further supports our recommendation in Section \ref{sec4} that portfolio managers consider the$1/x$ weighted portfolio.

We now consider the expected shortfall (cVaR) at 5\% for the eight portfolios under consideration.

\begin{table}[H]
\noindent \begin{centering}
\begin{tabular}{rccc}
\hline 
&Annual cVaR & Monthly cVaR & Daily cVaR \tabularnewline
\hline 
$1/x^{2}$ & -38.19  & -16.39  & -4.32 \tabularnewline
$1/x$ & -29.75  & -11.75  & -2.66 \tabularnewline
$1/\sqrt{x}$ & -28.28 & -10.94  & -2.45 \tabularnewline
$\log\left(x\right)$ & -27.09  & -10.33  & -2.36 \tabularnewline
EQU  & -26.90 & -10.10  & -2.34 \tabularnewline
$\sqrt{x}$ & -26.83  & -9.44  & -2.28 \tabularnewline
MKC  & -28.07  & -9.10  & -2.27 \tabularnewline
$x^{2}$ &  -29.23 & -9.10  & -2.41 \tabularnewline
\hline 
\end{tabular}
\par\end{centering}

\caption{Annual VaR, Monthly VaR, and Daily VaR for the EQU portfolio and the seven Tukey transformational ladder portfolios. All numbers are expressed in \%.} \label{table41}
\end{table}
As expected, the values for cVaR for the $1/x$ portfolio are much closer to the values of VaR for the EQU and MKC portfolios than the $1/x^2$ portfolio.

\section{Conclusion} \label{sec5}
Tukey's transformational ladder has proven to be a fundamental tool in modern data analysis, yet, to the best of our knowledge, has remained absent in its application to portfolio weighting. We have found that Tukey's transformational ladder remarkably produces several portfolios that obtain both cumulative and annual returns which exceed those of the traditional market capitalization weighted portfolio: they are the $1/x^2$, $1/x$, $1/\sqrt{x}$, $\text{log} (x)$, and $\sqrt{x}$ portfolios. Of these transformations, we have paid particular attention to $1/x^2$, as its average annual growth rate from 1958-2015 exceeds that of the market capitalization portfolio by approximately 12.11\%. However, due to both its large standard deviation and extremely high values of VaR and cVaR, the $1/x^2$ portfolio presents enormous risk, even for investors with long-term horizons. Investors should instead consider the $1/x$ weighted portfolio, which posts the highest Sharpe ratio of the eight portfolios under consideration, as well as more moderate values of VaR and cVaR than $1/x^2$.\\
\indent The current paper further raises a new and rather surprising phenomenon that both the cumulative and annual returns of our portfolios precisely follow the order of John Tukey's transformational ladder, exactly as it appeared in his seminal book on exploratory data analysis (\cite{Tukey2}):

\beqn
1/x^2,\,\, 1/x,\,\, 
1/\sqrt{x},\,\, \text{log}(x),\,\, \sqrt{x},\,\, x,\,\, x^2.
\eeqn
The theoretical foundation of the finding will be explored in a future paper.\\
\indent Finally, have noted that our empirical results neither contradict nor support the small-firm effect hypothesis, and therefore that results concerning Tukey's transformational ladder for portfolio management must be viewed as their own distinct phenomena.\\

\noindent \textbf{Acknowledgments}\\
We are very appreciative of an anonymous referee, whose helpful and detailed comments have enormously improved the quality of this work.

\newpage

\begin{center}
\textbf{Appendix A}\\ Tables \ref{table4} and \ref{table5} \label{append1}
\end{center}
\vspace{-.5cm}
\begin{table}[h!]
\begin{onehalfspace}
\noindent \begin{centering}
\begin{tabular}{rrrrrrrrr}

\hline 
Year  & $1/x^{2}$ & $1/x$ & $1/\sqrt{x}$ & $\log\left(x\right)$ & EQU  & $\sqrt{x}$ & MKC & $x^{2}$\tabularnewline
\hline 
1958 & 80.74 & 69.49 & 62.25 & 56.64 & 54.58 & 47.39 & 41.30 & 33.54 \tabularnewline
  1959 & 13.01 & 16.82 & 15.82 & 14.40 & 13.88 & 12.32 & 11.26 & 8.33 \tabularnewline
  1960 & -3.50 & -2.62 & -2.69 & -1.33 & -0.85 & -0.08 & -1.94 & -7.69 \tabularnewline
  1961 & 80.00 & 41.62 & 32.95 & 29.94 & 29.13 & 26.57 & 25.67 & 29.81 \tabularnewline
  1962 & -4.03 & -8.67 & -10.23 & -10.78 & -10.78 & -10.23 & -8.39 & -4.83 \tabularnewline
  1963 & 35.35 & 30.58 & 26.99 & 24.40 & 23.64 & 21.74 & 22.03 & 26.75 \tabularnewline
  1964 & 21.08 & 22.95 & 21.78 & 20.17 & 19.56 & 17.84 & 17.86 & 20.07 \tabularnewline
  1965 & 16.46 & 29.70 & 29.24 & 26.03 & 24.49 & 18.94 & 14.16 & 8.24 \tabularnewline
  1966 & -15.80 & -8.71 & -8.30 & -8.39 & -8.44 & -8.84 & -10.07 & -13.34 \tabularnewline
  1967 & 61.46 & 62.56 & 50.33 & 40.12 & 36.88 & 28.22 & 26.15 & 37.30 \tabularnewline
  1968 & 31.18 & 44.77 & 36.21 & 29.02 & 26.44 & 18.06 & 11.19 & 3.60 \tabularnewline
  1969 & -20.84 & -24.20 & -21.63 & -18.62 & -17.39 & -13.03 & -8.46 & 2.69 \tabularnewline
  1970 & 11.98 & 10.16 & 7.97 & 6.76 & 6.37 & 5.02 & 3.68 & -2.43 \tabularnewline
  1971 & 28.25 & 24.36 & 20.39 & 18.31 & 17.71 & 15.88 & 14.33 & 9.40 \tabularnewline
  1972 & 10.79 & 7.83 & 8.59 & 10.11 & 10.96 & 14.79 & 19.04 & 21.40 \tabularnewline
  1973 & -40.94 & -28.18 & -25.30 & -22.62 & -21.42 & -17.14 & -15.05 & -19.25 \tabularnewline
  1974 & -36.32 & -15.25 & -18.13 & -20.55 & -21.31 & -24.35 & -27.69 & -31.49 \tabularnewline
  1975 & 64.76 & 85.55 & 72.77 & 61.37 & 57.38 & 44.42 & 35.88 & 30.93 \tabularnewline
  1976 & 98.53 & 69.47 & 48.77 & 39.10 & 36.37 & 28.21 & 23.06 & 22.27 \tabularnewline
  1977 & 11.13 & 8.14 & 3.64 & -0.17 & -1.54 & -5.71 & -8.01 & -5.57 \tabularnewline
  1978 & 18.19 & 16.21 & 12.68 & 9.86 & 8.99 & 6.81 & 6.49 & 9.12 \tabularnewline
  1979 & 56.51 & 43.77 & 35.72 & 30.93 & 29.38 & 24.50 & 19.68 & 6.08 \tabularnewline
  1980 & 31.18 & 36.58 & 33.45 & 31.72 & 31.39 & 31.24 & 33.24 & 34.89 \tabularnewline
  1981 & 58.63 & 23.80 & 12.10 & 6.67 & 4.91 & -1.08 & -7.48 & -16.79 \tabularnewline
  1982 & 43.75 & 41.58 & 36.98 & 32.95 & 31.28 & 25.74 & 21.71 & 27.72 \tabularnewline
  1983 & 56.16 & 46.34 & 38.45 & 33.09 & 31.21 & 25.67 & 22.29 & 24.17 \tabularnewline
  1984 & 14.45 & 5.29 & 3.59 & 3.51 & 3.69 & 4.51 & 5.55 & 6.45 \tabularnewline
  1985 & 8.84 & 21.89 & 28.18 & 31.07 & 31.69 & 32.68 & 31.97 & 29.35 \tabularnewline
  1986 & 50.50 & 26.18 & 19.88 & 18.61 & 18.65 & 18.99 & 17.95 & 2.86 \tabularnewline
  1987 & 27.10 & 15.03 & 10.15 & 8.40 & 7.94 & 6.55 & 5.58 & 3.91 \tabularnewline
  1988 & 38.25 & 33.24 & 26.76 & 23.16 & 22.07 & 18.95 & 16.70 & 12.95 \tabularnewline
  \hline 
\end{tabular}\caption{Annual returns (in \%) for the EQU and the seven Tukey transformational ladder portfolios from 1958 to 1988.}\label{table4}

\par\end{centering}
\end{onehalfspace}

\end{table}

\begin{table}
\begin{onehalfspace}
\noindent \centering{}%
\begin{tabular}{rrrrrrrrr}
\hline 
Year  & $1/x^{2}$ & $1/x$ & $1/\sqrt{x}$ & $\log\left(x\right)$ & EQU  & $\sqrt{x}$ & MKC & $x^{2}$\tabularnewline
\hline 
  1989 & -27.53 & 10.95 & 21.92 & 25.85 & 26.91 & 29.87 & 31.24 & 23.94 \tabularnewline
  1990 & 28.14 & -9.61 & -14.94 & -12.51 & -11.22 & -6.88 & -2.77 & 3.70 \tabularnewline
  1991 & 126.86 & 68.96 & 46.24 & 38.67 & 37.12 & 33.01 & 30.29 & 27.27 \tabularnewline
  1992 & -3.49 & 12.46 & 16.57 & 16.37 & 15.62 & 12.10 & 7.81 & 1.65 \tabularnewline
  1993 & 12.35 & 18.15 & 17.86 & 16.45 & 15.70 & 12.97 & 10.25 & 5.81 \tabularnewline
  1994 & -3.19 & 1.41 & 1.87 & 1.71 & 1.61 & 1.38 & 1.53 & 2.03 \tabularnewline
  1995 & -8.88 & 21.75 & 28.85 & 31.86 & 32.83 & 35.80 & 38.08 & 39.82 \tabularnewline
  1996 & 15.91 & 18.48 & 19.36 & 20.03 & 20.42 & 22.15 & 24.81 & 30.82 \tabularnewline
  1997 & 52.93 & 30.93 & 28.38 & 28.77 & 29.40 & 32.00 & 34.41 & 35.73 \tabularnewline
  1998 & 2.88 & 5.97 & 9.00 & 12.22 & 13.96 & 21.16 & 29.43 & 40.21 \tabularnewline
  1999 & -22.10 & 2.62 & 9.22 & 11.52 & 12.36 & 16.23 & 22.05 & 32.73 \tabularnewline
  2000 & -37.31 & -3.01 & 9.18 & 11.78 & 10.91 & 3.79 & -7.29 & -24.06 \tabularnewline
  2001 & 6.30 & 18.55 & 11.15 & 4.38 & 1.72 & -6.71 & -11.82 & -13.37 \tabularnewline
  2002 & 98.77 & 1.17 & -11.84 & -15.63 & -16.50 & -19.06 & -21.25 & -25.18 \tabularnewline
  2003 & 101.63 & 64.94 & 51.28 & 44.44 & 42.17 & 34.83 & 28.41 & 20.43 \tabularnewline
  2004 & 29.22 & 21.44 & 19.44 & 18.21 & 17.56 & 14.67 & 10.85 & 6.63 \tabularnewline
  2005 & -20.56 & 0.82 & 5.82 & 7.71 & 7.95 & 7.50 & 5.06 & 0.31 \tabularnewline
  2006 & -5.07 & 13.84 & 16.19 & 16.43 & 16.35 & 15.91 & 15.68 & 16.43 \tabularnewline
  2007 & -7.40 & -5.19 & -2.44 & -0.10 & 0.86 & 3.78 & 5.60 & 7.17 \tabularnewline
  2008 & -33.54 & -36.87 & -37.90 & -38.09 & -37.96 & -37.07 & -35.25 & -31.01 \tabularnewline
  2009 & 128.19 & 82.03 & 63.41 & 52.59 & 48.92 & 37.42 & 27.59 & 12.37 \tabularnewline
  2010 & 30.43 & 27.08 & 24.93 & 23.15 & 22.28 & 18.98 & 15.49 & 10.90 \tabularnewline
  2011 & -1.44 & -1.11 & -0.39 & 0.07 & 0.24 & 0.82 & 1.83 & 5.88 \tabularnewline
  2012 & 26.32 & 20.39 & 18.61 & 17.73 & 17.47 & 16.72 & 16.02 & 14.05 \tabularnewline
  2013 & 43.06 & 39.40 & 37.67 & 36.67 & 36.29 & 34.78 & 32.17 & 23.56 \tabularnewline
  2014 & 15.65 & 14.01 & 14.18 & 14.41 & 14.47 & 14.36 & 13.70 & 13.20 \tabularnewline
  2015 & -7.52 & -5.29 & -3.92 & -2.87 & -2.36 & -0.50 & 1.46 & 3.10 \tabularnewline
  \hline 
Arithmetic  & 23.92 & 20.35 & 17.40 & 15.62 & 15.03 & 13.18 & 11.81 & 10.25\tabularnewline
Geometric & 18.00 & 17.53 & 15.23 & 13.80 & 13.32 & 11.73 & 10.43 & 8.69\tabularnewline
SD & 39.54 & 26.44 & 22.29 & 20.01 & 19.30 & 17.52 & 16.98 & 18.05\tabularnewline
VaR (annual) & -33.96 & -16.60 & -18.65 & -18.91 & -17.98 & -17.43 & -15.98 & -24.23\tabularnewline
\hline 
\end{tabular}\caption{Annual returns (in \%) for the EQU and the seven Tukey transformational ladder portfolios from 1988 to 2015. The arithmetic means, geometric means, and standard deviations, and annual VaR of each portfolio, calculated over 1958-2015 and inclusive of all dividends and transaction fees, are also displayed.} \label{table5}
\end{onehalfspace}
\end{table}

\newpage

\begin{center}
\textbf{APPENDIX B} \\ Why rebalance monthly?
 \end{center}

\begin{figure}[H]
\begin{centering}
\includegraphics[width=16.6cm]{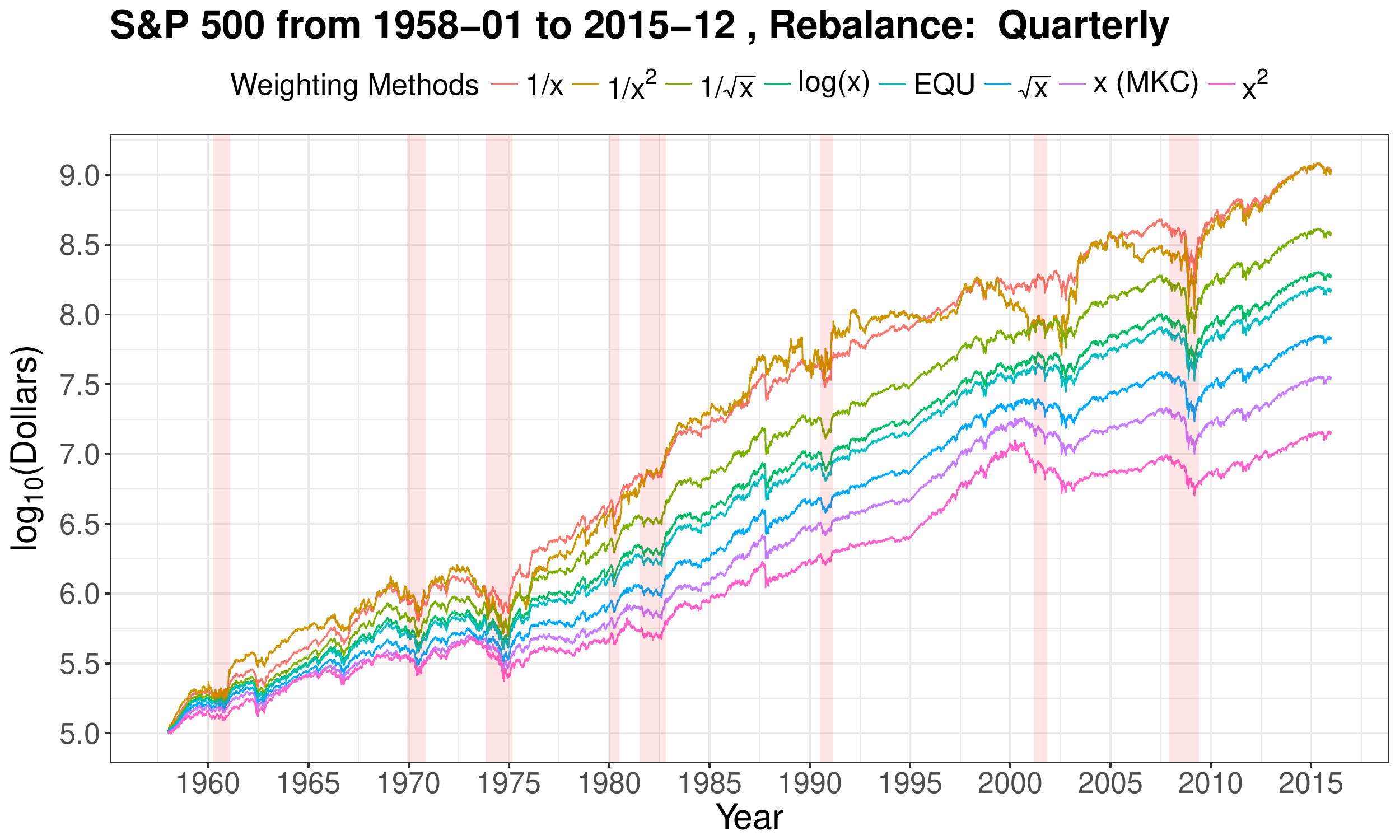}
\par\end{centering}
\caption{Tukey transformational ladder with quarterly rebalancing from 1958 to
2015.} \label{fig20}
\end{figure}

In this Appendix, we show that it is advantageous for investors holding the $1/x^2$ portfolio to rebalance their portfolios \textit{monthly}.\\ 
\indent In all of the calculations below, we begin with \$100,000 in 1958 dollars. We assume transaction administrative fees of \$1 (in 2015 dollars) per trade and, additionally, a long-run average bid-ask spread of .1\% of the closing value of the stock. Rebalancing daily, the portfolio goes broke. Having already considered monthly balancing shown as Figure \ref{fig3} in the main document, we now turn to an analysis of quarterly rebalancing and yearly rebalancing. \\
\indent We first consider quarterly rebalancing. Figure \ref{fig20} displays the cumulative returns calculated from 1958 to 2015 of the equally weighted S\&P 500 portfolio (EQU) and the seven Tukey transformational ladder portfolios ($1/x^2,\,\, 1/x,\,\, 1/\sqrt{x},\,\, \text{log}(x),\,\, \sqrt{x},\,\,x,\,\,x^2$), where $x$ is the market capitalization weighted portfolio, and the portfolios are rebalanced quarterly. 

\noindent The cumulative returns displayed in Figure \ref{fig20} are reproduced in Table \ref{table50} below.
\begin{table}[H]
\footnotesize
\centering{}%
\begin{tabular}{rrrrrrrr}
\hline 
$1/x^{2}$ & $1/x$  & $1/\sqrt{x}$  & $\log\left(x\right)$  & EQU  & $\sqrt{x}$  & $x$  & $x^{2}$\tabularnewline
\hline 
\$1.054 bil  & \$1.081 bil & \$377.268 mil & \$187.874 mil  & \$148.360 mil  & \$67.326 mil  & \$34.959 mil  & \$14.113 mil \tabularnewline
\hline 
\end{tabular}\caption{Ending balance on 12/31/15.} \label{table50}
\end{table}

\newpage
We next consider annual rebalancing. Figure \ref{fig21} below displays the cumulative returns calculated from 1958 to 2015 of the equally weighted S\&P 500 portfolio (EQU) and the seven portfolios given by the Tukey transformations ($1/x^2,\,\, 1/x,\,\, 1/\sqrt{x},\,\, \text{log}(x),\,\, \sqrt{x},\,\,x,\,\,x^2$), where $x$ is the market capitalization weighted portfolio, and the portfolios are rebalanced annually.

\begin{figure}[H]
\begin{centering}
\includegraphics[width=16.6cm]{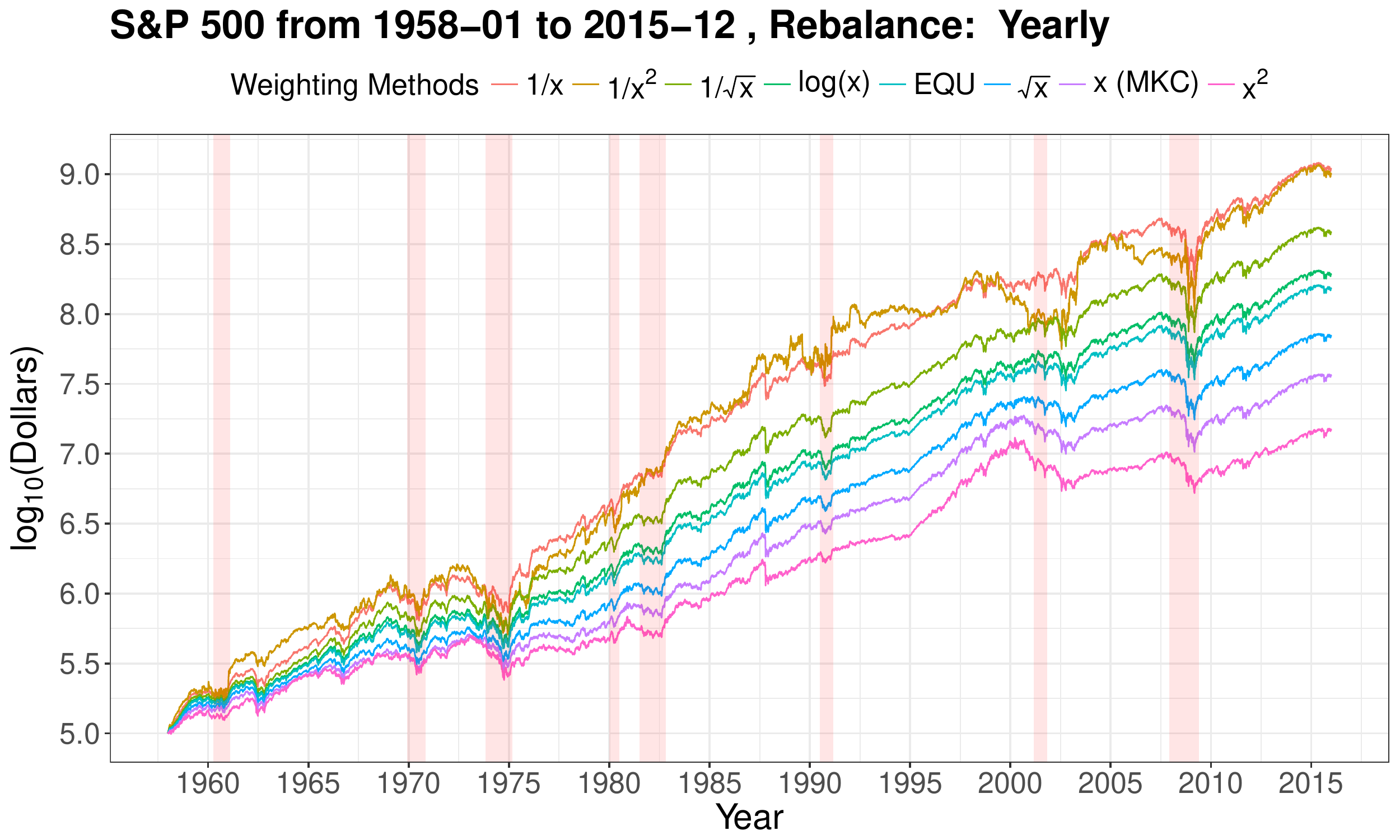}
\par\end{centering}
\caption{Tukey transformation ladder with yearly rebalancing from 1958 to 2015.} \label{fig21}
\end{figure}
\noindent The cumulative returns displayed in Figure \ref{fig21} are reproduced in Table \ref{table51} below.
\begin{table}[H]
\footnotesize
\centering{}%
\begin{tabular}{rrrrrrrr}
\hline 
$1/x^{2}$ & $1/x$  & $1/\sqrt{x}$  & $\log\left(x\right)$  & EQU  & $\sqrt{x}$  & $x$  & $x^{2}$\tabularnewline
\hline 
\$999.798 mil  & \$1.084 bil  & \$381.412 mil & \$190.942 mil & \$151.005 mil  & \$69.160 mil  & \$36.106 mil  & \$14.692 mil \tabularnewline
\hline 
\end{tabular}\caption{Ending balance on 12/31/15.} \label{table51}
\end{table}

We conclude by summarizing the findings of Figures \ref{fig20} and \ref{fig21} for the $1/x^2$ portfolio. When rebalanced quarterly, the balance of the $1/x^2$ portfolio on 12/31/15 is \$1.054 billion. When rebalanced annually, the value of the $1/x^2$ portfolio on 12/31/15 \$999.798 million. The \$1.477 billion figure for the ending balance on 12/31/15 for the monthly rebalanced $1/x^2$ portfolio (Table \ref{table2}) exceeds that of both quarterly rebalancing (Table \ref{table50}) and of annual rebalancing (Table \ref{table51}).

\newpage

\begin{center}
\textbf{APPENDIX C} \\ Results of Bootstrap for Random $N$
\end{center}

This Appendix displays the bootstrapped distributions for fixed $N$ for a seven different values of $N$\\($N=20,50,100,200,300,400,500$). The results herein are presented to support our findings in Section \ref{sec62} of the main manuscript.\\\\
\textbf{C.1: $N=20$}

\begin{figure}[H]
\noindent \centering{}\includegraphics[width=16cm]{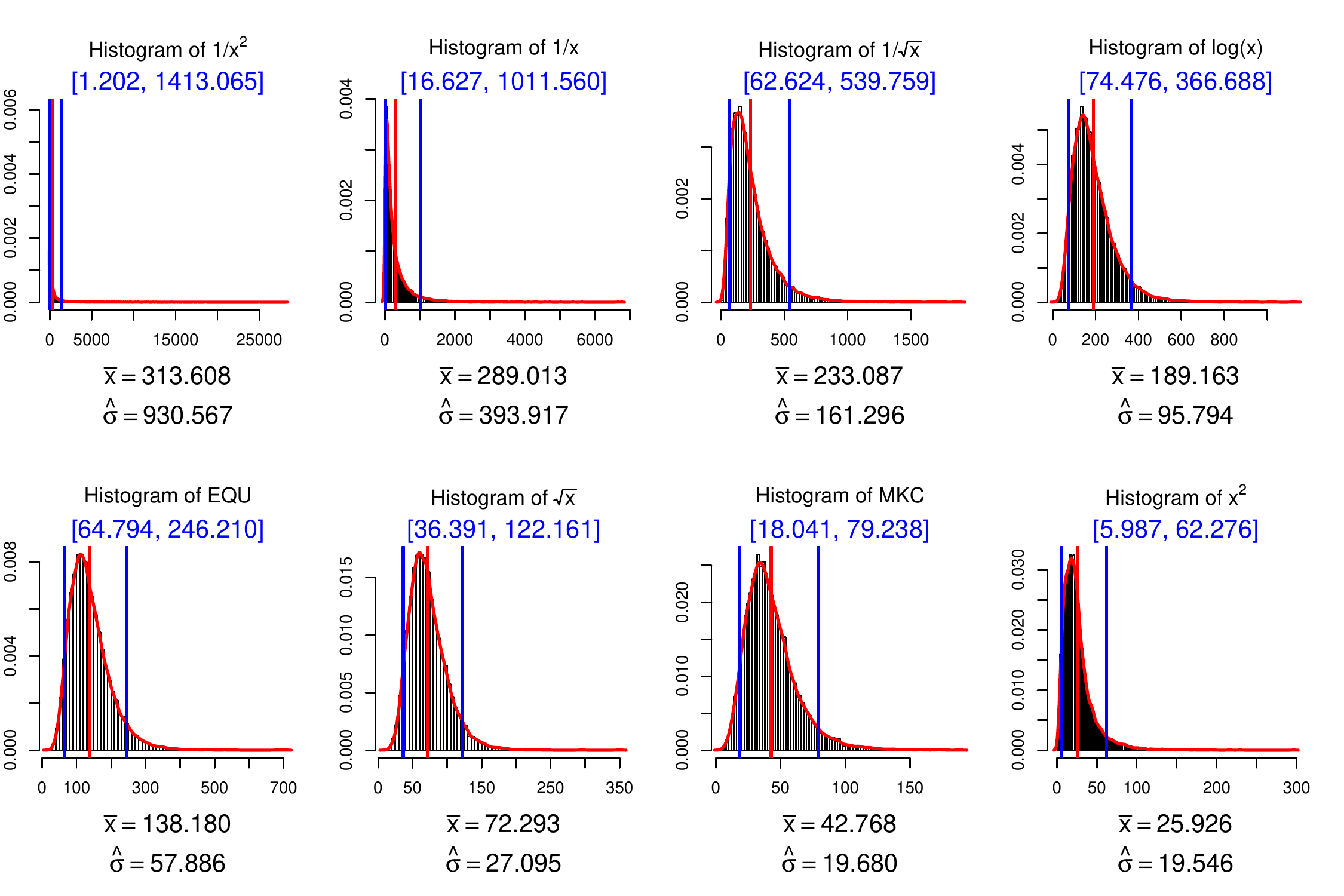}\caption{The bootstrap distribution for $N=20$.} 
\end{figure}

\begin{table}[H]
\begin{onehalfspace}
\noindent \begin{centering}
\begin{tabular}{rrrrrrrr}
\hline 
 & mean  & median  & sd  & q5th  & q95th  & q1th & q99th\tabularnewline
\hline 
$1/x^{2}$ & 313.608 & 53.904 & 930.567 & 1.202 & 1413.065 & 0.206 & 4117.225 \tabularnewline 
$1/x$ & 289.013 & 156.083 & 393.917 & 16.627 & 1011.560 & 6.456 & 1903.231 \tabularnewline 
$1/\sqrt{x}$ & 233.087 & 192.686 & 161.296 & 62.624 & 539.759 & 38.731 & 805.649 \tabularnewline 
 $\log\left(x\right)$ & 189.163 & 169.573 & 95.794 & 74.476 & 366.688 & 53.044 & 509.182 \tabularnewline 
EQU & 138.180 & 127.712 & 57.886 & 64.794 & 246.210 & 48.370 & 322.390 \tabularnewline 
$\sqrt{x}$ & 72.293 & 68.113 & 27.095 & 36.391 & 122.161 & 27.708 & 155.877 \tabularnewline 
$x$ & 42.768 & 39.249 & 19.680 & 18.041 & 79.238 & 12.319 & 108.881 \tabularnewline 
$x^{2}$ & 25.926 & 21.423 & 19.546 & 5.987 & 62.276 & 3.236 & 100.276 \tabularnewline 
\hline 
\end{tabular}
\par\end{centering}
\end{onehalfspace}

\caption{Sample statistics for the cumulative return on 12/31/15 for $N=20$, calculated from 20,000 simulations. All numbers are in million USD.}
\end{table}

\newpage

\textbf{C.2: $N=50$}

\begin{figure}[H]
\noindent \centering{}\includegraphics[width=16cm]{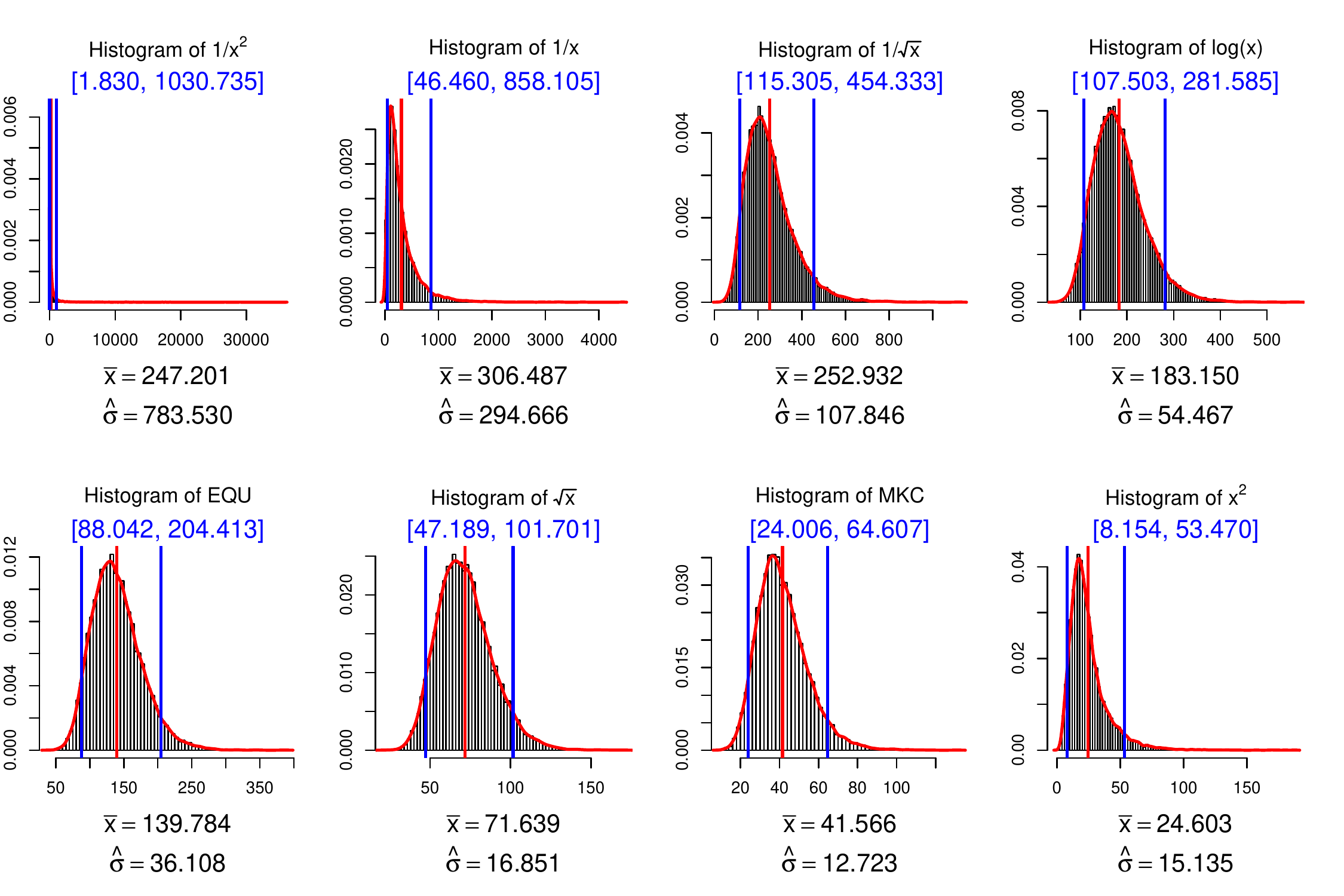}\caption{The bootstrap distribution for $N=50$.}
\end{figure}

\begin{table}[H]
\begin{onehalfspace}
\noindent \begin{centering}
\begin{tabular}{rrrrrrrr}
\hline 
 & mean  & median  & sd  & q5th  & q95th  & q1th & q99th\tabularnewline
\hline 
$1/x^{2}$ & 247.201 & 51.832 & 783.530 & 1.830 & 1030.735 & 0.476 & 3127.920 \tabularnewline 
$1/x$ & 306.487 & 216.857 & 294.666 & 46.460 & 858.105 & 22.401 & 1444.515 \tabularnewline 
$1/\sqrt{x}$ & 252.932 & 233.828 & 107.846 & 115.305 & 454.333 & 84.473 & 592.379 \tabularnewline 
 $\log\left(x\right)$ & 183.150 & 176.354 & 54.467 & 107.503 & 281.585 & 86.982 & 342.475 \tabularnewline 
EQU & 139.784 & 135.746 & 36.108 & 88.042 & 204.413 & 73.265 & 243.520 \tabularnewline 
$\sqrt{x}$ & 71.639 & 70.064 & 16.851 & 47.189 & 101.701 & 39.665 & 118.217 \tabularnewline 
$x$ & 41.566 & 39.805 & 12.723 & 24.006 & 64.607 & 18.968 & 79.438 \tabularnewline 
$x^{2}$ & 24.603 & 20.977 & 15.135 & 8.154 & 53.470 & 5.084 & 79.391 \tabularnewline 
\hline 
\end{tabular}
\par\end{centering}
\end{onehalfspace}

\caption{Sample statistics for the cumulative return on 12/31/15 for $N=50$, calculated from 20,000 simulations. All numbers are in million USD.}
\end{table}

\newpage

\textbf{C.3: $N=100$}

\begin{figure}[H]
\noindent \centering{}\includegraphics[width=16cm]{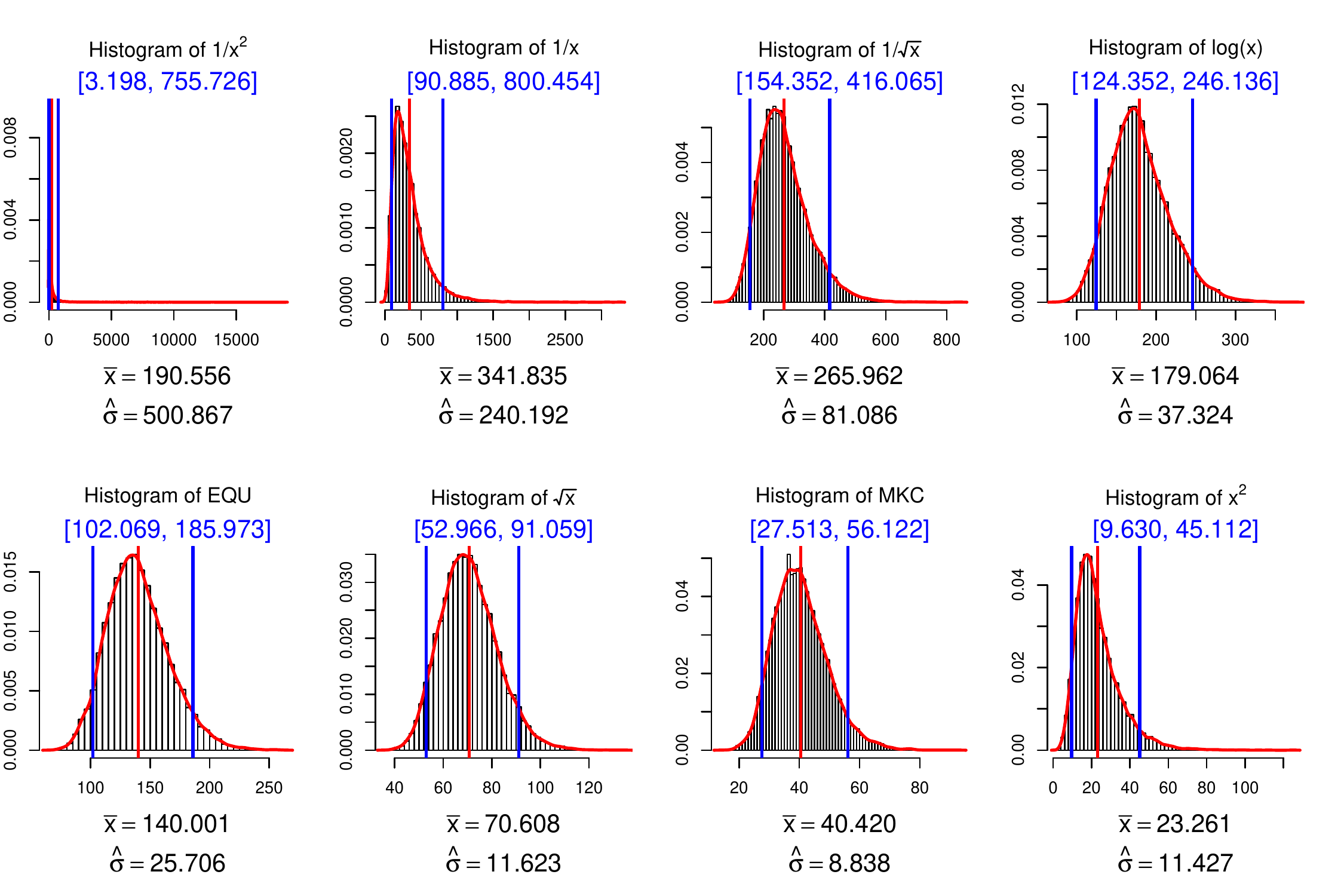}\caption{The bootstrap distribution for $N=100$.}
\end{figure}

\begin{table}[H]
\begin{onehalfspace}
\noindent \begin{centering}
\begin{tabular}{rrrrrrrr}
\hline 
 & mean  & median  & sd  & q5th  & q95th  & q1th & q99th\tabularnewline
\hline 
$1/x^{2}$ & 190.556 & 55.298 & 500.867 & 3.198 & 755.726 & 0.902 & 2128.342 \tabularnewline 
$1/x$ & 341.835 & 282.757 & 240.192 & 90.885 & 800.454 & 53.339 & 1196.739 \tabularnewline 
$1/\sqrt{x}$ & 265.962 & 254.665 & 81.086 & 154.352 & 416.065 & 123.279 & 503.724 \tabularnewline 
 $\log\left(x\right)$ & 179.064 & 175.209 & 37.324 & 124.352 & 246.136 & 107.527 & 281.018 \tabularnewline 
EQU & 140.001 & 137.717 & 25.706 & 102.069 & 185.973 & 89.326 & 209.102 \tabularnewline 
$\sqrt{x}$ & 70.608 & 69.817 & 11.623 & 52.966 & 91.059 & 47.388 & 101.690 \tabularnewline 
$x$ & 40.420 & 39.613 & 8.838 & 27.513 & 56.122 & 23.357 & 65.239 \tabularnewline 
$x^{2}$ & 23.261 & 20.712 & 11.427 & 9.630 & 45.112 & 6.529 & 61.034 \tabularnewline 
\hline 
\end{tabular}
\par\end{centering}
\end{onehalfspace}

\caption{Sample statistics for the cumulative return on 12/31/15 for $N=100$, calculated from 20,000 simulations. All numbers are in million USD.}
\end{table}

\newpage

\textbf{C.4: $N=200$}

\begin{figure}[H]
\noindent \centering{}\includegraphics[width=16cm]{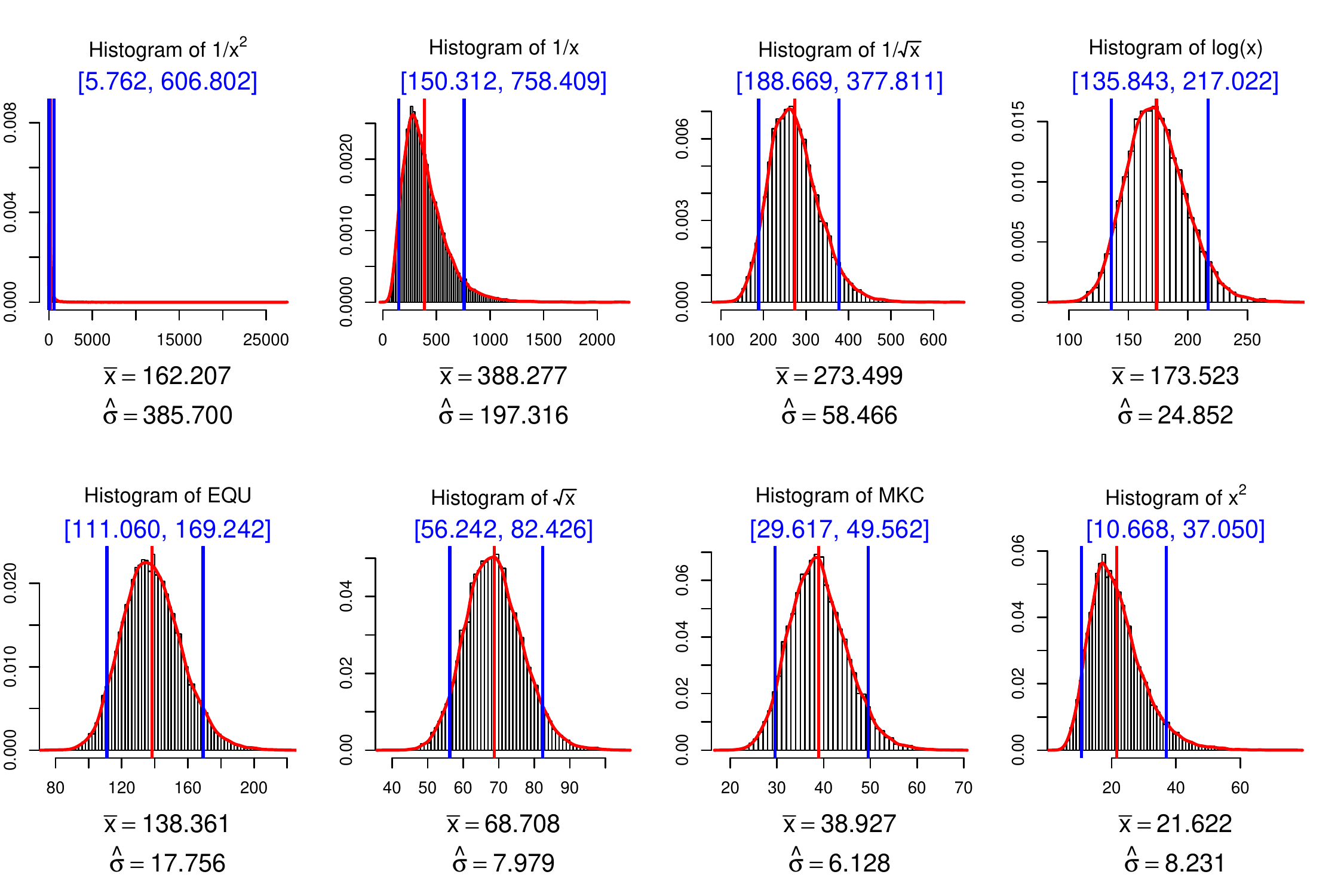}\caption{The bootstrap distribution for $N=200$.}
\end{figure}

\begin{table}[H]
\begin{onehalfspace}
\noindent \begin{centering}
\begin{tabular}{rrrrrrrr}
\hline 
 & mean  & median  & sd  & q5th  & q95th  & q1th & q99th\tabularnewline
\hline 
$1/x^{2}$ & 162.207 & 64.156 & 385.700 & 5.762 & 606.802 & 2.036 & 1508.764 \tabularnewline 
$1/x$ & 388.277 & 347.982 & 197.316 & 150.312 & 758.409 & 105.384 & 1039.324  \tabularnewline  
$1/\sqrt{x}$ & 273.499 & 267.685 & 58.466 & 188.669 & 377.811 & 162.252 & 434.943 \tabularnewline  
 $\log\left(x\right)$ & 173.523 & 171.956 & 24.852 & 135.843 & 217.022 & 122.656 & 238.693 \tabularnewline 
EQU & 138.361 & 137.392 & 17.756 & 111.060 & 169.242 & 101.419 & 184.743 \tabularnewline 
$\sqrt{x}$ & 68.708 & 68.347 & 7.979 & 56.242 & 82.426 & 51.664 & 89.029 \tabularnewline 
$x$ & 38.927 & 38.550 & 6.128 & 29.617 & 49.562 & 26.363 & 54.955 \tabularnewline 
$x^{2}$ & 21.622 & 20.345 & 8.231 & 10.668 & 37.050 & 7.854 & 46.815 \tabularnewline 
\hline 
\end{tabular}
\par\end{centering}
\end{onehalfspace}
\caption{Sample statistics for the cumulative return on 12/31/15 for $N=200$, calculated from 20,000 simulations. All numbers are in million USD.}
\end{table}

\newpage

\textbf{C.5: $N=300$}

\begin{figure}[H]
\noindent \centering{}\includegraphics[width=16cm]{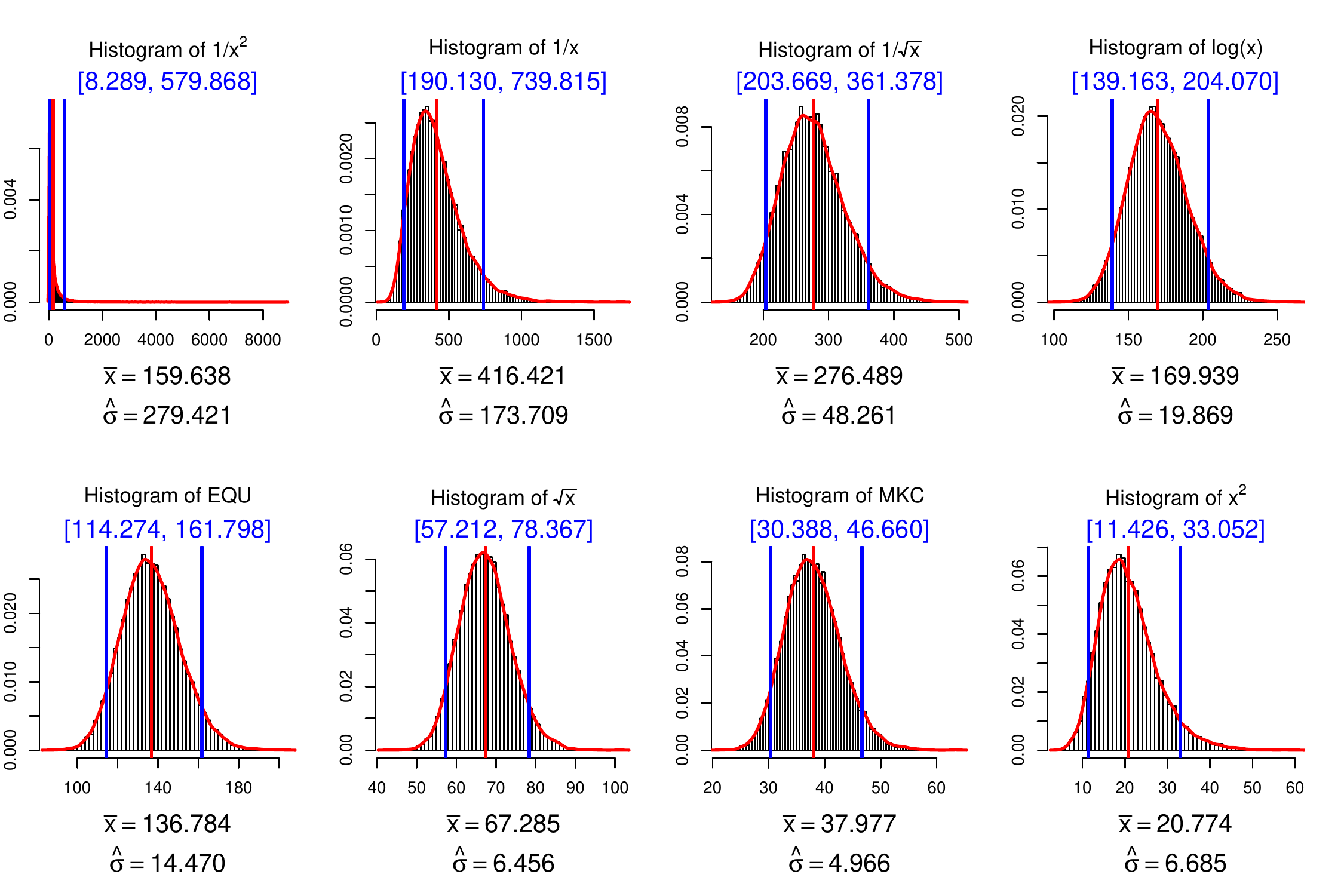}\caption{The bootstrap distribution for $N=300$.}
\end{figure}

\begin{table}[H]
\begin{onehalfspace}
\noindent \begin{centering}
\begin{tabular}{rrrrrrrr}
\hline 
 & mean  & median  & sd  & q5th  & q95th  & q1th & q99th\tabularnewline
\hline 
$1/x^{2}$ & 159.638 & 73.879 & 279.421 & 8.289 & 579.868 & 3.437 & 1324.234 \tabularnewline 
$1/x$ & 416.421 & 386.993 & 173.709 & 190.130 & 739.815 & 140.194 & 959.616 \tabularnewline 
$1/\sqrt{x}$ & 276.489 & 272.826 & 48.261 & 203.669 & 361.378 & 181.529 & 405.815 \tabularnewline 
 $\log\left(x\right)$ & 169.939 & 168.763 & 19.869 & 139.163 & 204.070 & 128.248 & 220.758 \tabularnewline 
EQU & 136.784 & 136.063 & 14.470 & 114.274 & 161.798 & 106.171 & 173.457 \tabularnewline 
$\sqrt{x}$ & 67.285 & 67.017 & 6.456 & 57.212 & 78.367 & 53.540 & 83.961 \tabularnewline 
$x$ & 37.977 & 37.692 & 4.966 & 30.388 & 46.660 & 27.711 & 50.785 \tabularnewline 
$x^{2}$ & 20.774 & 19.895 & 6.685 & 11.426 & 33.052 & 8.649 & 40.593 \tabularnewline 
\hline 
\end{tabular}
\par\end{centering}
\end{onehalfspace}

\caption{Sample statistics for the cumulative return on 12/31/15 for $N=300$, calculated from 20,000 simulations. All numbers are in million USD.}
\end{table}

\newpage

\textbf{C.6: $N=400$}

\begin{figure}[H]
\noindent \centering{}\includegraphics[width=16cm]{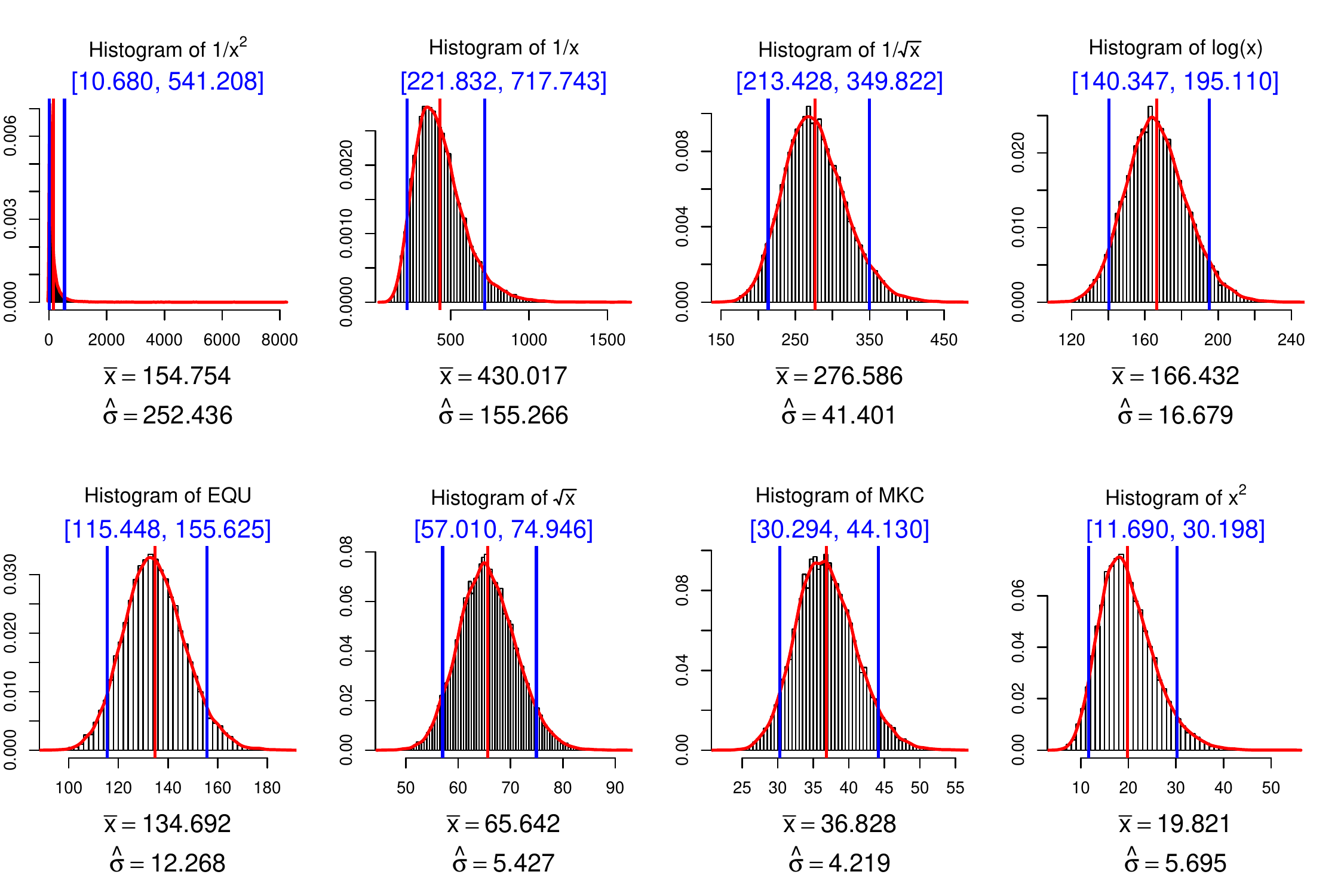}\caption{The bootstrap distribution for $N=400$. }
\end{figure}

\begin{table}[H]
\begin{onehalfspace}
\noindent \begin{centering}
\begin{tabular}{rrrrrrrr}
\hline 
 & mean  & median  & sd  & q5th  & q95th  & q1th & q99th\tabularnewline
\hline 
$1/x^{2}$ & 154.754 & 78.818 & 252.436 & 10.680 & 541.208 & 4.452 & 1185.654 \tabularnewline 
$1/x$ & 430.017 & 406.943 & 155.266 & 221.832 & 717.743 & 172.303 & 901.348 \tabularnewline 
$1/\sqrt{x}$ & 276.586 & 273.788 & 41.401 & 213.428 & 349.822 & 192.470 & 385.304 \tabularnewline 
 $\log\left(x\right)$ & 166.432 & 165.592 & 16.679 & 140.347 & 195.110 & 130.664 & 208.605 \tabularnewline 
EQU & 134.692 & 134.124 & 12.268 & 115.448 & 155.625 & 108.059 & 165.356 \tabularnewline 
$\sqrt{x}$ & 65.642 & 65.399 & 5.427 & 57.010 & 74.946 & 54.037 & 79.065 \tabularnewline 
$x$ & 36.828 & 36.594 & 4.219 & 30.294 & 44.130 & 27.915 & 47.650 \tabularnewline 
$x^{2}$ & 19.821 & 19.133 & 5.695 & 11.690 & 30.198 & 9.375 & 35.838 \tabularnewline 
\hline 
\end{tabular}
\par\end{centering}
\end{onehalfspace}

\caption{Sample statistics for the cumulative return  on 12/31/15 for $N=400$, calculated from 20,000 simulations. All numbers are in million USD.}
\end{table}

\newpage

\textbf{C.7: $N=500$}

\begin{figure}[H]
\noindent \centering{}\includegraphics[width=16cm]{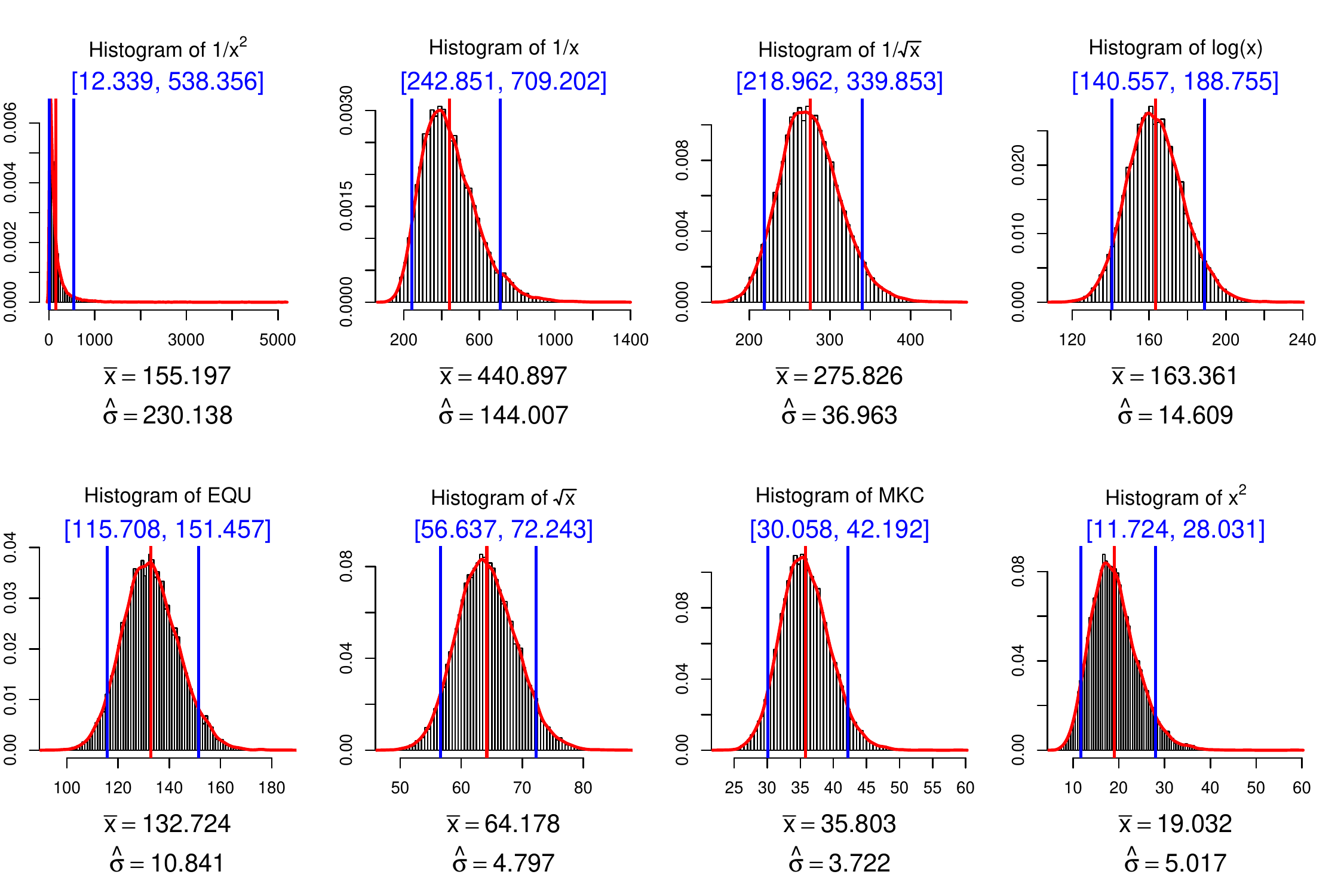}\caption{The bootstrap distribution for $N=500$.}
\end{figure}

\begin{table}[H]
\begin{onehalfspace}
\noindent \begin{centering}
\begin{tabular}{rrrrrrrr}
\hline 
 & mean  & median  & sd  & q5th  & q95th  & q1th & q99th\tabularnewline
\hline 
$1/x^{2}$ & 155.197 & 83.330 & 230.138 & 12.339 & 538.356 & 5.825 & 1057.348 \tabularnewline
$1/x$ & 440.897 & 420.618 & 144.007 & 242.851 & 709.202 & 189.991 & 860.461 \tabularnewline
$1/\sqrt{x}$ & 275.826 & 273.563 & 36.963 & 218.962 & 339.853 & 199.851 & 372.607 \tabularnewline
 $\log\left(x\right)$ & 163.361 & 162.654 & 14.609 & 140.557 & 188.755 & 131.992 & 199.961 \tabularnewline
EQU & 132.724 & 132.275 & 10.841 & 115.708 & 151.457 & 109.453 & 159.636 \tabularnewline
$\sqrt{x}$ & 64.178 & 63.957 & 4.797 & 56.637 & 72.243 & 53.787 & 76.126 \tabularnewline
$x$ & 35.803 & 35.583 & 3.722 & 30.058 & 42.192 & 27.859 & 45.267 \tabularnewline
$x^{2}$ & 19.032 & 18.518 & 5.017 & 11.724 & 28.031 & 9.607 & 33.065 \tabularnewline
\hline 
\end{tabular}
\par\end{centering}
\end{onehalfspace}

\caption{Sample statistics for the cumulative return  on 12/31/15 for $N=500$, calculated from 20,000 simulations. All numbers are in million USD.}
\end{table}

\newpage

\begin{center}
\textbf{APPENDIX D} \\ Value of S\&P 500 Portfolio over the 1980-2015, 1990-2015, and 2000-2015 horizons
\end{center}

In this Appendix, we show that the returns of the eight portfolios under consideration precisely follow the order of the Tukey transformational ladder for three additional time horizons: 1980-2015, 1990-2015, and 2000-2015. \\
\indent We first consider the 1980-2015 horizon. We invest \$272,028 on 1/2/80 (the equivalent of \$100,000 in 1958 dollars) and let each portfolio grow until 12/31/15. The cumulative returns are displayed below.

\begin{figure}[H] \label{fig88}
\noindent \begin{centering}
\includegraphics[width=16.6cm]{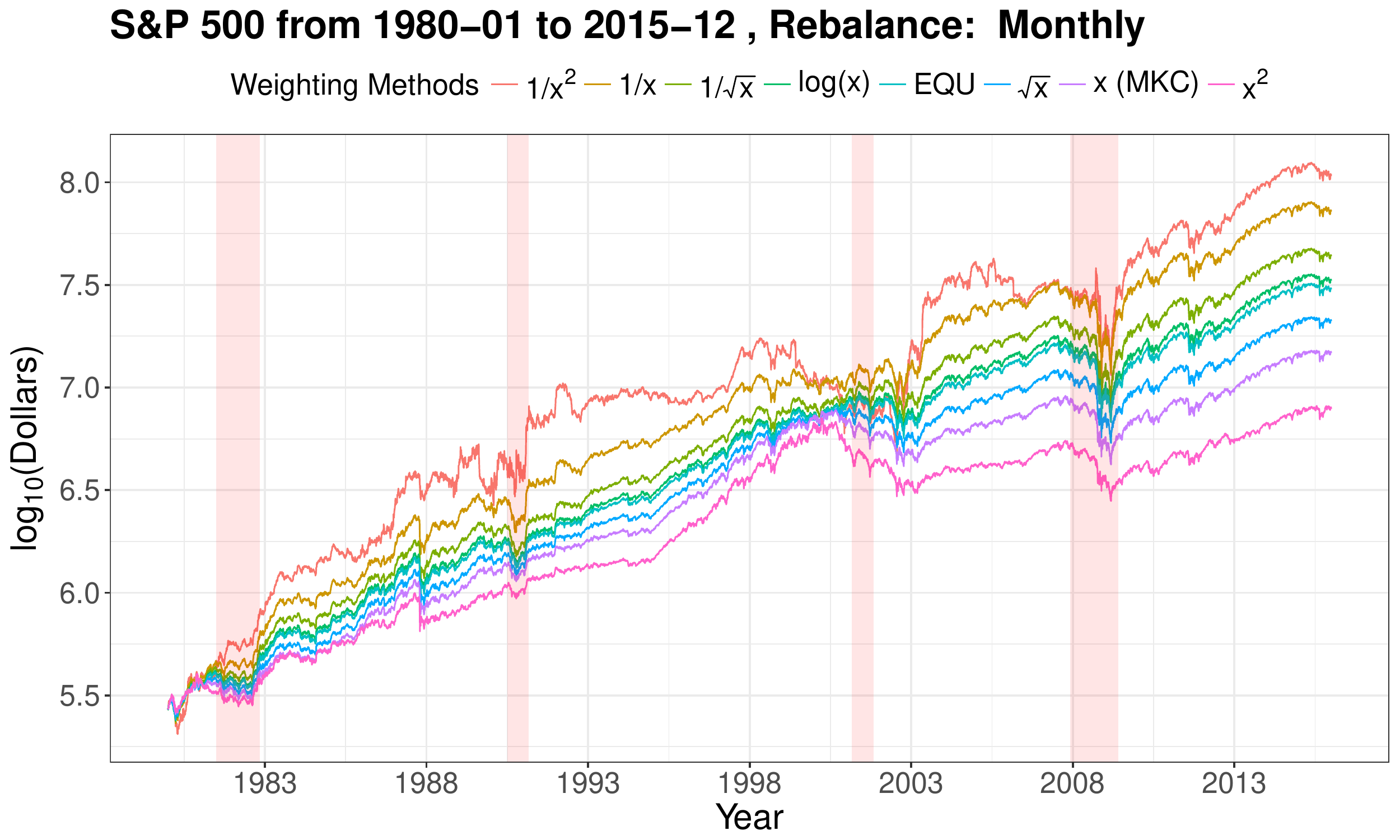} 
\par\end{centering}

\caption{Cumulative $\text{log}_{10}$ returns (from 1980-2015) for the EQU portfolio and the seven Tukey transformational ladder portfolios. The calculation assumes that \$272,028 is invested on 1/2/80 and is left to grow until 12/31/15.}
\end{figure}

\noindent The cumulative returns on 12/31/15 displayed in the above Figure are reproduced in Table \ref{table10} below.

\begin{table}[H]
\begin{onehalfspace}
\footnotesize
\noindent \centering{}%
\begin{tabular}{rrrrrrrr}
\hline 
$1/x^{2}$ & $1/x$ & $\sqrt{x}$ & $\log\left(x\right)$  & EQU & $\sqrt{x}$ & $x$ & $x^{2}$\tabularnewline
\hline 
\$107.967 mil  & \$72.113 mil  & \$ 43.637 mil & \$ 33.160 mil & \$ 30.086 mil  & \$ 21.066 mil  & \$14.730 mil  & \$ 7.875 mil \tabularnewline
\hline 
\end{tabular}\caption{The cumulative returns for the EQU portfolio and the seven Tukey transformational ladder portfolios. The calculation assumes that \$272,028 is invested on 1/2/80 and is left to grow until 12/31/15.} \label{table10}
\end{onehalfspace}
\end{table}

\newpage
We now consider the 1990-2015 time horizon. We invest \$445,455 on 1/2/90 (the equivalent of \$100,000 in 1958 dollars) and let the portfolios grow until 12/31/15. The results are displayed in Figure \ref{fig11} below.

\begin{figure}[H]
\noindent \begin{centering}
\includegraphics[width=16.6cm]{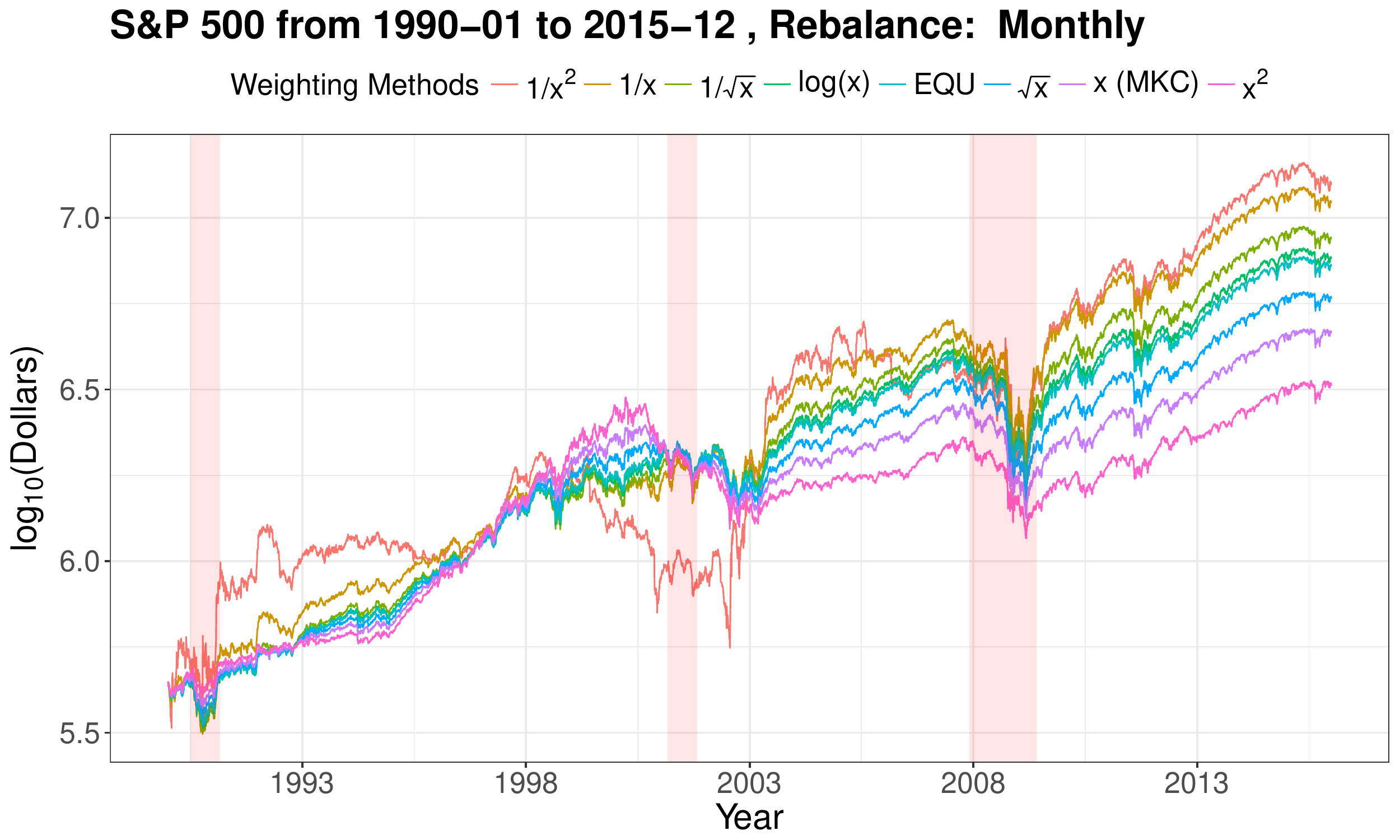}
\par\end{centering}

\caption{Cumulative $\text{log}_{10}$ returns (from 1990-2015) for the EQU portfolio and the seven Tukey transformational ladder portfolios. The calculation assumes that \$445,455 is invested on 1/2/90 and is left to grow until 12/31/15.} \label{fig11}
\end{figure}

\noindent The cumulative returns displayed in Figure \ref{fig11} are reproduced in Table \ref{table11} below

\begin{table}[H]
\begin{onehalfspace}
\footnotesize
\noindent \centering{}%
\begin{tabular}{rrrrrrrr}
\hline 
$1/x^{2}$ & $1/x$ & $\sqrt{x}$ & $\log\left(x\right)$  & EQU & $\sqrt{x}$ & $x$ & $x^{2}$\tabularnewline
\hline 
\$12.527 mil  & \$11.041 mil & \$8.646 mil  &\$7.582 mil  & \$7.197 mil & \$5.808 mil  & \$4.608 mil & \$3.242 mil \tabularnewline
\hline
\end{tabular}\caption{The cumulative returns for the EQU portfolio and the seven Tukey transformational ladder portfolios. The calculation assumes that \$445,455 is invested on 1/2/90 and is left to grow until 12/31/15.} \label{table11}
\end{onehalfspace}
\end{table}

\newpage

Finally, we consider the 2000-2015 time horizon. We invest \$590,210 on 1/2/00 (the equivalent of \$100,000 in 1958 dollars) and let the portfolios grow until 12/31/15. We display the results in Figure \ref{fig12} below.

\begin{figure}[H]
\noindent \begin{centering}
\includegraphics[width=16.6cm]{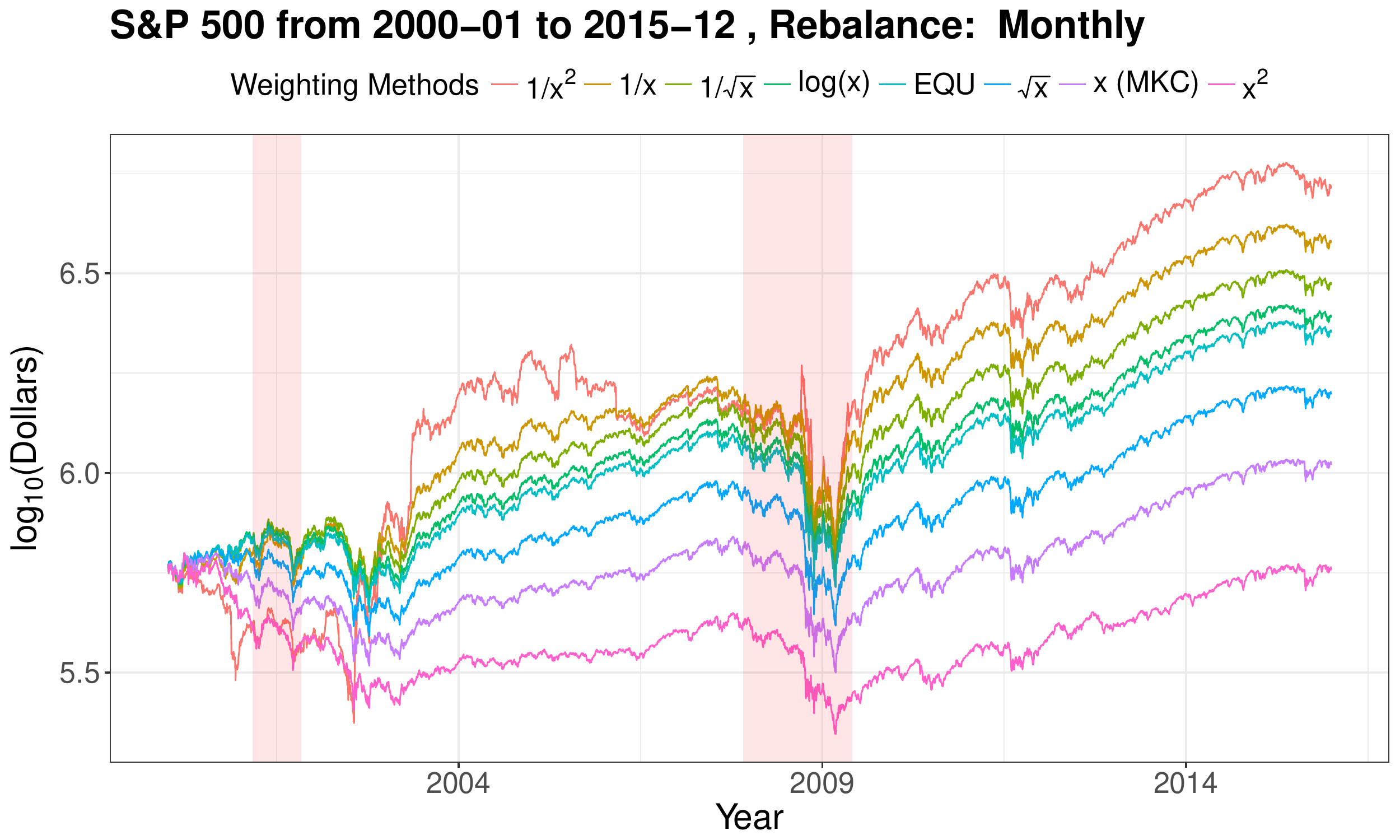}
\par\end{centering}

\caption{Cumulative $\text{log}_{10}$ returns (from 2000-2015) for the EQU portfolio and the Tukey transformational ladder portfolios. The calculation assumes that \$590,210 is invested on 1/2/00 and is left to grow until 12/31/15.} \label{fig12}
\end{figure}
\noindent The cumulative returns displayed in Figure \ref{fig12} are produced in Table \ref{table12} below
\begin{table}[H]
\begin{onehalfspace}
\footnotesize
\noindent \centering{}%
\begin{tabular}{rrrrrrrr}
\hline 
$1/x^{2}$ & $1/x$ & $\sqrt{x}$ & $\log\left(x\right)$  & EQU & $\sqrt{x}$ & $x$ & $x^{2}$\tabularnewline
\hline 
\$5.176 mil  & \$3.762 mil  & \$2.952 mil  & \$2.446 mil & \$2.243 mil  & \$ 1.572 mil & \$1.048 mil  & \$.570 mil \tabularnewline
\hline 
\end{tabular}\caption{The cumulative returns for the EQU portfolio and the seven Tukey transformational ladder portfolios. The calculation assumes that \$590,210 is invested on 1/2/00 and is left to grow until 12/31/15.} \label{table12}
\end{onehalfspace}
\end{table}
In conclusion, tables \ref{table10}, \ref{table11}, \ref{table12} each show that the portfolio returns precisely follow the order of the Tukey transformational ladder. 
\newpage

\bibliographystyle{apalike}
\bibliography{VF}

\noindent \textbf{Supplementary Materials}
For purposes of replication, all code used in this work can be found online on the following GitHub repository: \url{https://github.com/yinsenm/Tukeytransforms}. 

\newpage
\noindent \textbf{Short CVs}\\\\
\textbf{Philip Ernst} is an Assistant Professor at Rice University’s Department of Statistics. He received his Ph.D. in 2014 from The Wharton School of the University of Pennsylvania and joined the Rice faculty the same year. He holds a M.A. in Statistics from The Wharton School of the University of Pennsylvania and a B.A. in Statistics \textit{cum laude} from Harvard University. His research interests are mathematical finance, applied probability, and statistical theory. In the past three years, he has published four articles in mathematical finance (\cite{Ernst}, \cite{Ernst1}, \cite{Ernst2}, \cite{Ernst3}).

\noindent \textbf{James R. Thompson} is Emeritus Noah Harding Professor and Research Professor of Statisitics at Rice University. He holds a BE in Chemical Engineering from Vanderbilt and a doctorate in Mathematics from Princeton. He is the author or co-author of 14 books and over 100 articles. A Fellow of the ASA, the IMS and the ISI, he is recipient
of the Army’s Wilks Award for contributions related to National Defense. He has held adjunct professor rank at the M.D. Anderson Cancer Center and the UTS School of Public Health. His primary current interests are in epidemiology, oncology, business, and quality control.

\noindent \textbf{Yinsen Miao} is a second year Ph.D. student in the Department of Statistics at Rice University. His research interests include portfolio optimization, statistical machine learning, convex optimization, and Bayesian models. His research focuses on developing statistical algorithms for sparse data. He received his M.A. in Statistics from Rice University and his bachelor degree from Minzu University of China.

\end{document}